\documentclass[%
reprint,
pra,
]{revtex4-1}

\usepackage[utf8x]{inputenc}
\usepackage{graphicx}

\usepackage{amsmath}
\usepackage{amssymb}
\usepackage{siunitx}
\usepackage{mathtools}

\usepackage{hyperref}
\usepackage[pdftex,dvipsnames]{xcolor}
\hypersetup{frenchlinks = true,
	colorlinks,
	linkcolor={red!50!black},
	citecolor={blue!50!black},
	urlcolor={blue}
}

\usepackage{lipsum}
\usepackage{braket}
\usepackage{enumitem}

\usepackage{lmodern}
\usepackage{slantsc}

\usepackage{xargs}                      
\usepackage[colorinlistoftodos,prependcaption,]{todonotes}
\newcommandx{\unsure}[2][1=]{\todo[linecolor=red,backgroundcolor=red!25,bordercolor=red,#1]{#2}}
\newcommandx{\change}[2][1=]{\todo[linecolor=blue,backgroundcolor=blue!25,bordercolor=blue,#1]{#2}}
\newcommandx{\info}[2][1=]{\todo[linecolor=OliveGreen,backgroundcolor=OliveGreen!25,bordercolor=OliveGreen,#1]{#2}}
\newcommandx{\improvement}[2][1=]{\todo[linecolor=Plum,backgroundcolor=Plum!25,bordercolor=Plum,#1]{#2}}
\newcommandx{\thiswillnotshow}[2][1=]{\todo[disable,#1]{#2}}

\paperwidth=\dimexpr \paperwidth + 6cm\relax 
\oddsidemargin=\dimexpr\oddsidemargin + 3cm\relax
\evensidemargin=\dimexpr\evensidemargin + 3cm\relax
\marginparwidth=\dimexpr \marginparwidth + 3cm\relax

\def\op#1{\hat{#1}}
\def\st#1{_{\mathrm{#1}}}
\def\ut#1{^{\mathrm{#1}}}

\renewcommand{\d}[0]{\mathrm{d}}

\usepackage{xspace}
\newcommand{\bec}{\textsc{bec}\xspace}
\newcommand{\grape}{\textsc{grape}\xspace}
\newcommand{\pgrape}{\textsc{pgrape}\xspace}
\newcommand{\rs}{\textsc{rs}\xspace}

\newcommand{\pr}{\textsc{p}r\xspace}
\newcommand{\sa}{\textsc{sa}\xspace}

\newcommand{\ps}{\textsc{ps}\xspace}
\newcommand{\po}{\textsc{po}\xspace}

\newcommand{\dbscan}{\textsc{dbscan}\xspace}
\newcommand{\psitgt}[1][]{%
	\ifthenelse{\equal{#1}{}}{\psi_{\mathrm{tgt}}}{\psi_{\mathrm{tgt,#1}}}%
}
\newcommand{\U}[0]{\op{\mathcal{U}}}

\begin{document}

	\title{Crowdsourcing human common sense for quantum control 
	}
	\author{Jesper Hasseriis Mohr Jensen}
	\author{Miroslav Gajdacz}
	\author{Shaeema Zaman Ahmed}
	\author{Jakub Herman Czarkowski}
	\author{Carrie Weidner}
	\author{Janet Rafner}
	\author{Jens Jakob S\o rensen}
	\author{Klaus M\o lmer}	
	\author{Jacob Friis Sherson}	
	\email{sherson@phys.au.dk}
	\author{\textit{Quantum Moves 2} players}	
	
	\affiliation{%
		Department of Physics and Astronomy, Aarhus University, Ny Munkegade 120, 8000 Aarhus C, Denmark
	}%

	\date{\today}
	
	\begin{abstract}
\if
Citizen science methodologies have over the past decade been applied with great success to help solve highly complex numerical challenges within the field of microbiology, e.g. protein folding. Such citizen science interfaces are currently being explored as creative support tools for domain experts and as benchmarking for state-of-the-art machine learning methods. In principle, quantum physics offers similarly complex challenges, yet there does not exist a systematic research effort exploring the potential of citizen science in this domain. Here, we take first steps in this direction by introducing a citizen science game, \textit{Quantum Moves 2}, and across three different quantum optimal control problems  compare the performance of different optimization methods: gradient-based local optimization with player-based seeds and uniform random seeds, as well as gradient-free (discrete coordinate) stochastic ascent with uniform random seeds. 
Although approximately optimal results can be obtained within reasonable computational resources, 
the three considered problems fall into distinct categories: i) \textit{Splitting} has a very simple optimization landscape with a single broad attractor and associated solution strategy, ii) \textit{Bring Home Water} has two distinct solution strategies characterized by an exponentially-widening fidelity gap, and some methods exhibit difficulties locating the optimal one, and iii) \textit{Shake Up} has a more complex structure of intermixed strategies, and within the allotted resources for optimization we cannot claim to have located the optimum using any of the methods. 
Inside the game, players can apply the gradient-based algorithm (running locally on the device) to their seeds and we
find that these results perform roughly on par with the best of the standard optimization methods performed on a computer cluster.
This highlights the future potential for crowdsourcing the solution of quantum research problems. 
In addition, cluster-optimized player seeds was the only method to exhibit roughly optimal performance throughout all three challenges. We are aware that making comparisons with entirely random seeding constitutes the lowest possible bar of comparison with the plethora of possible numerical optimization approaches and our work should therefore by no means be taken as more than a necessary first demonstration of the potential for further exploration. Additionally, since the three challenges are significantly simpler than the problem of folding proteins, they also lend themselves to different research questions. Where protein folding focuses on the results of a small subset of players achieving near expert status, we explore the bulk behavior of all players. The relative success of the player seeds indicates, in our view, the potential value of studying the crowdsourcing of common sense, the innate responses of all players to the interface, as inspiration for expert optimization.
\fi

\if11
Citizen science methodologies have over the past decade been applied with great success to help solve highly complex numerical challenges. Here, we take early steps in the quantum physics arena by introducing a citizen science game, \textit{Quantum Moves 2}, and 
compare the performance of different optimization methods across three different quantum optimal control problems of varying difficulty.
Inside the game, players can apply a gradient-based algorithm (running locally on their device) to optimize their solutions and we
find that these results perform roughly on par with the best of the tested standard optimization methods performed on a computer cluster.
In addition, cluster-optimized player seeds was the only method to exhibit roughly optimal performance across all three challenges.
This highlights the potential for crowdsourcing the solution of future quantum research problems. 
\fi
	\end{abstract}

	\maketitle
	


\section{Introduction}
\label{sec:introduction}

Despite amazing advances in the past years, it has becoming increasingly clear that pattern-matching results from deep learning algorithms alone can be surprisingly brittle \cite{heaven2019deep,jiang2019vulnerability}. This failure has been attributed to, among other reasons, a lack of hierarchical learning, transfer between problems, and common sense comprehension of real world phenomena \cite{marcus2020next}. Common sense has been defined in many different ways, but here we refer to it as implicit and shared fundamental assumptions that people have about the world \cite{fletcher1986psychology}. Additionally, there are indications that humans may sometimes solve computationally hard problems quickly and near-optimally \cite{carruthers2013evaluating,koepnick2019novo}. However, humans just as often fail miserably \cite{kahneman2011thinking}. Thus, many argue that optimized synergetic systems integrating individuals or collectives of humans and machines offer a promising, human-in-the-loop, approach to tackle complex problems \cite{dellermann2019hybrid, michelucci2016power,kamar2016directions, berditch2020future,baltz2017achievement}. One key challenge in this approach is that it requires large-scale studies of human capacities, for example common sense and the development of rich cognitive models \cite{marcus2020next}. Initial steps in this direction can be taken by exploring problems in research-relevant contexts, such as in the related fields of citizen science and collective intelligence \cite{michelucci2016power,berditch2020future}, and detailed comparisons between human  \cite{koepnick2019novo} and AI \cite{senior2020improved} performance are becoming feasible for problems such as protein folding.

\begin{figure}[t]
	
			\includegraphics[width=1.1\linewidth]{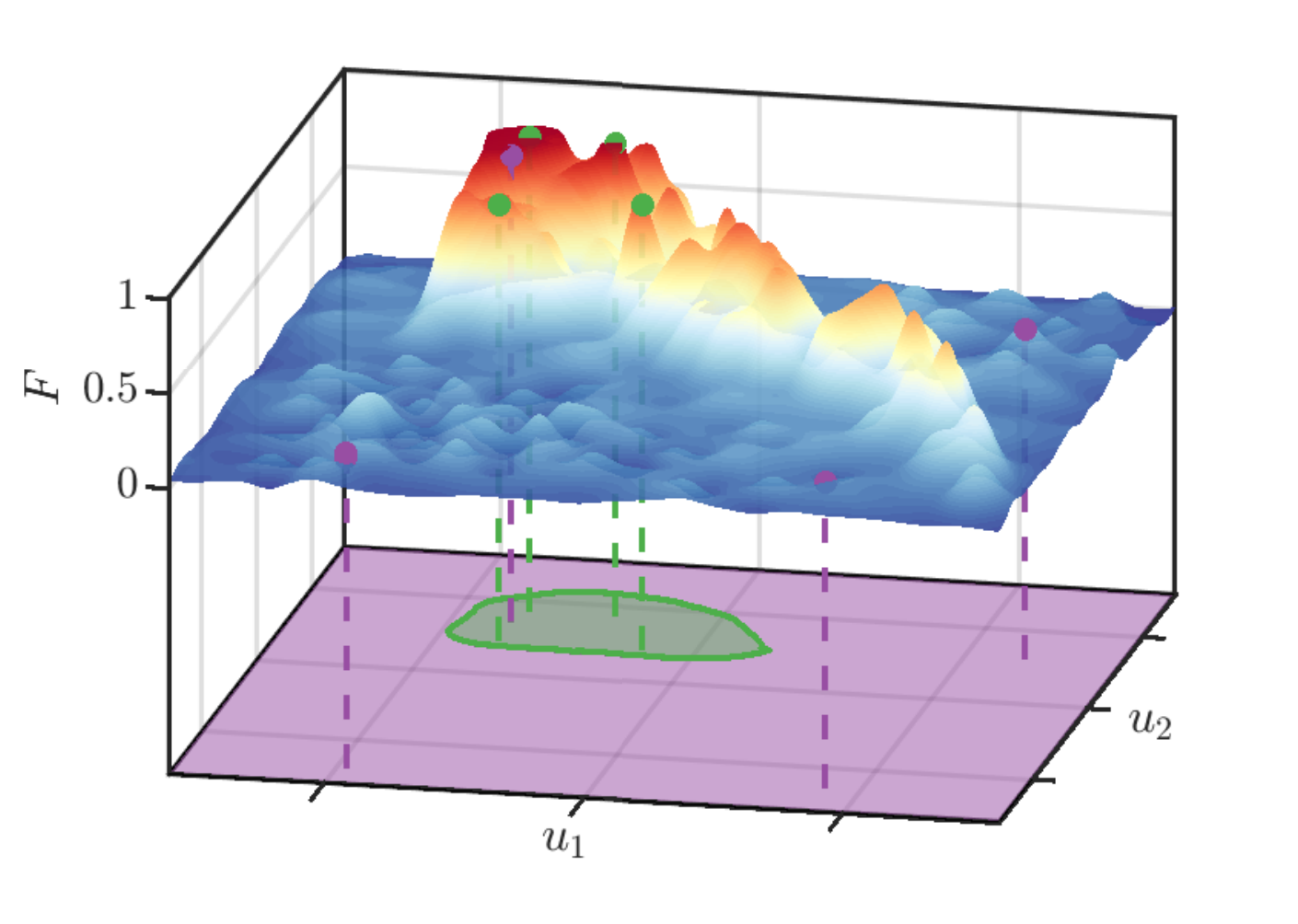}
	\caption{Abstract illustration of an optimization (control) landscape with two control parameters. 
		Each point in the plane corresponds to a particular set of functions $\{u_1(t), u_2(t)\}$ and the height of the landscape is given by the associated fidelity.		
		Only a small region of the control space is optimal (with respect to $F\approx 1$) and otherwise contains many local ``traps'' -- the relative efficiency of a local ``hillclimbing'' optimizer is thus correlated to where it is seeded.
		Purple region: Uniform random seeding. Obtaining $F\approx 1$ is possible, but with low probability.
		Green region: Specialized seeding. Obtaining $F\approx 1$ is possible with high probability, but not guaranteed.
		Such a region is conventionally targeted by experts through their accumulated knowledge, problem insights, heuristics, and programming proficiency. Non-expert citizen scientist players may be useful in uncovering this region without extensive prior training.}
	\label{fig:landscape}
\end{figure}

\medskip
In this paper, we take steps towards exploring how citizen input can be applied to the solution of relevant problems in quantum mechanics. 
In particular, we consider a dynamic quantum state transfer task (elaborated further in the following section) where the goal is to maximize the \textit{fidelity}, $0 \leq F \leq 1$, by finding appropriate control functions $\{u_1(t),u_2(t)\}$ (solutions) as illustrated in Fig.~\ref{fig:landscape}. These are typically identified by locally ascending \textit{seeds} (starting points) to the nearest maximum in the associated optimization landscape. Possible seeding strategies range from uniform random guessing (least imposed structure) to parameterizations based on highly domain-specific expert knowledge or heuristics (most imposed structure) as depicted by the colored regions in Fig.~\ref{fig:landscape}, each with associated probabilities of obtaining $F\approx 1$ upon optimization. Discovering the ``good'' regions of the optimization landscape can often be very challenging.
In this work, we investigate whether non-expert citizen scientists may be useful in efficiently identifying good regions by gathering their input through our game, \textit{Quantum Moves 2}, which allows the player to both create seeds and subsequently engage with a local optimization algorithm embedded in the game and see its action in real time. 
Each player can be considered as an independent, adaptive seeding strategy that incorporates high-level heuristics and complex decision-making processes into an optimization loop which are otherwise difficult to capture and implement programmatically. 
This methodology may be used as a general means to extract features and heuristics for a given problem 
which could then aid experts in further analysis and guide the development of seeding strategies.
We analyze three distinct scientific challenges -- i.e. optimization landscapes -- where the performance of the hybrid approach (human-computer) is compared against standard methodologies from quantum optimal control and computer science. 
In this sense, this work does not aim to present or promote an inimical competition between players and computer algorithms, but rather, explore their possible interplay in terms of solution strategy and the usefulness of player seeding in such problems.
Our optimization code is available at \url{https://gitlab.com/quatomic/quantum-moves2}.

Note that a simpler game interface, \textit{Quantum Moves 1}, was previously accessible, and studies of the user experience \cite{lieberoth2015getting,diaz2020more} and numerics of the \textit{Bring Home Water} level \cite{sorensen2016exploring} were published.
However, due to an error in the optimization code \cite{ gronlund2020explaining}, the quantitative comparisons presented in the publication Ref.~\cite{sorensen2016exploring} are not valid and the article has been retracted. The research presented in the present work, which is broader both in terms of scope and analyses, was conducted with a wholly new codebase, including algorithms explored in \cite{sels2018stochastic,gronlund2019algorithms}, and free of the error that hampered the results in Ref.~\cite{sorensen2016exploring}.

\medskip
In order to address the tenability of game-based exploration of quantum research problems, we pose two specific questions about the current and potential scientific contribution to quantum physics and related fields:

\begin{enumerate}[label=Q\arabic*:]
	\item  Can a suitable gamified interface allow citizen scientists to solve quantum control problems entirely on their own, using their own hardware to run the required computation (e.g. using an in-game optimizer), with a quality on par with traditional expert-driven optimization? 
	\item Can the player-generated solutions, in combination with concrete algorithms, provide an edge against fully algorithmic solutions, and how does that depend on the type and mathematical complexity of the problem? 
\end{enumerate}

If Q1 can be answered in the affirmative, then such an interface would allow for a novel form of online, quantum citizen science combining human problem solving with crowd computing. In such a framework, one could imagine quantum researchers continually feeding in optimization challenges that are then solved efficiently by the community at no computational cost to the researchers.

To address these questions in a systematic manner and to establish a baseline, we compare optimization of player seeding to uniform random seeding across a range of distinct problems.
As we shall see below, for the investigated problems, player seeds are on average more efficient than the randomly-generated ones. 
This broad contribution of the players stands in stark contrast to e.g. citizen science projects like Foldit \cite{cooper2010predicting}, where only a small fraction of players provide a scientific contribution after an extensive training process. In order to distinguish these two types of citizen science challenges, we assert in this work that the \textit{Quantum Moves 2} game, to some extent, taps into certain aspects of common sense (shared tacit knowledge of reasonable behavior) of the player population at large. Although it is not within the scope of this work to analyze the explicit nature of this common sense, it seems reasonable to conjecture that the liquid analogy of the wave function dynamics taps into the classical intuition for sloshing water, which was also argued by D. Sels in his numerical analysis of one of the problems \cite{sels2018stochastic}.

\section{Quantum Optimal Control}
\label{sec:qoc}
In this section, we describe the context and define key goals of \textit{Quantum Moves 2}.

The hallmarks of the second quantum revolution \cite{dowling2003quantum} are the exploitation and engineering of fragile, isolated quantum objects. Quantum computing with any platform predicates precise control of the constituent qubits and associated gate operations. 
Additionally, the control must also be expeditiously carried out such as to avoid decoherence and other detrimental effects to the overall goal. Controls meeting these criteria and more can be obtained within the well-established theory of quantum optimal control. 

A common class of quantum optimal control problems deals with facilitating a particular initial-to-target state transfer, $\ket{\psi_0} \rightarrow \ket{\psitgt}$, for some fixed process duration $T$. 
The manipulatory access to the state evolution is through a set of \textit{control} parameters $\{u_1(t),u_2(t),\dots\}$
where each solution (specific choice of functions $u_i(t)$) uniquely maps to a final state $\ket{\psi_0} \rightarrow \ket{\psi(T)}$ 
(see Appendices~\ref{app:numerics}-\ref{app:algorithms}). 
In this context, the transfer fidelity 
\begin{align}
0 \leq F[\{u_1(t),u_2(t),\dots\}; T]=|\braket{\psi\st{tgt}|\psi(T)}|^2 \leq 1,
\end{align}
for each fixed $T$ can be interpreted as a high-dimensional optimization (or control) landscape as illustrated in Fig.~\ref{fig:landscape}.
\textit{Optimal} controls (or solutions) can then be associated with points in the landscape that are globally or locally maximal.
Additionally, for a given fidelity requirement, we associate a fundamental quantum speed limit, $T^{F}\st{QSL}$, below which no maximum exceeds $F$.
Thus, 
$T\st{QSL}^{F}$ is defined as the shortest duration at which at least one control corresponding to a maximum can obtain the given $F$. 
Depending on the context, common choices for threshold values are $F=0.99,0.999,0.9999, \dots$,
characterized by a trade-off between $F$ and $T\st{QSL}^{F}$ (increased precision leads to longer durations).

In the limit $T\rightarrow \infty$, most problems become easy in the sense that many global maxima with $F\approx 1$ exist. As $T \rightarrow 0$, however, the control problem becomes increasingly difficult as 
previously global maxima gradually become only locally maximal and the control landscape becomes more rugged.
The usually unfavorable topography of the control landscape in the $T\approx T^{F}\st{QSL}$ regime therefore makes uncovering global maxima especially difficult for the aforementioned high-fidelity requirements. 

At its core, any iterative optimization algorithm attempting to locate global maxima must prescribe a way to traverse the optimization landscape in a meaningful way. It must thus strike a balance between local (exploitation) and global (exploration) search methodologies.
A common optimization paradigm initializes an algorithm, e.g. one excelling in finding the nearest local optimum, from many different seeds, thereby introducing a simple global component \cite{ugray2007multistart}. 
In principle, a local optimizer maps each seed to its nearest \textit{attractor} \footnote{Numerical and implementation details such as inexact line searches introduce the possibility of ``skipping over'' the true nearest attractor}. 
Non-global local optima are often called local \textit{traps}, referring to the propensity for said optimizers to locate these and terminate (since they are ``stuck''). 
The effectiveness of this paradigm is then necessarily strongly correlated to a combination of 
the seeding strategy (the mechanism with which seeds are generated) and the choice of optimization algorithm. 

Effective seeding strategies, e.g. those targeting the green region of optimality depicted in Fig.~\ref{fig:landscape}, and algorithms naturally become increasingly important with growing problem 
complexity and this can broadly be characterized by two axes: the computational (or numerical) complexity of the underlying simulations and the inherent topographical landscape complexity. 
The computational difficulty can also be interpreted as the amount of required resources.
Below, we discuss problems with two different degrees of computational difficulty drawing upon the single particle Schr\"odinger equation and the non-linear dynamics of Bose-Einstein condensates.



\section{Overview of Quantum Moves 2}
\label{sec:levels}
In \textit{Quantum Moves 2}, the player's goal is to solve various state transfer problems, referred to in-game as \textit{levels}. 
Each level concerns 1D transfers of either single particle or Bose-Einstein condensate (\bec) wave functions $\psi(x,t) = \braket{x|\psi(t)}$, both describable by the Hamiltonian
\begin{align}
\op H\st{} &= -\frac{\hbar^2}{2m} \frac{\partial^2 }{\partial x^2} + V + g |\psi|^2,
\label{eq:H}
\end{align}
where taking the non-linear coupling parameter $g=0$ corresponds to the single-particle case. 
While there is no in-game distinction in their representation, 
the numerics involving \bec ($g\neq 0$) are more intricate, and this has consequences for the efficiency of some optimization algorithms as discussed later.
The potential has up to two controllable parameters, $V = V(u_1(t),u_2(t))$, depending on the level. 

The main features of \textit{Quantum Moves 2} are 
\begin{enumerate}
   \setlength\itemsep{0.1mm}
	\item A device-embedded algorithm that enables players optimize the seeds they produce.
	\item In-game tools for analyzing previous solutions.
	\item Constrained exploration to the regime near $T\st{QSL}^{F}$ and below with $F=0.99,0.999$.
	\item Scientific problems of various complexity.
\end{enumerate}
\pgfmathsetmacro{\thewidth}{0.73}
\begin{figure}
		\includegraphics[width=\thewidth\linewidth, trim={10cm 0cm 10cm 1cm},clip]{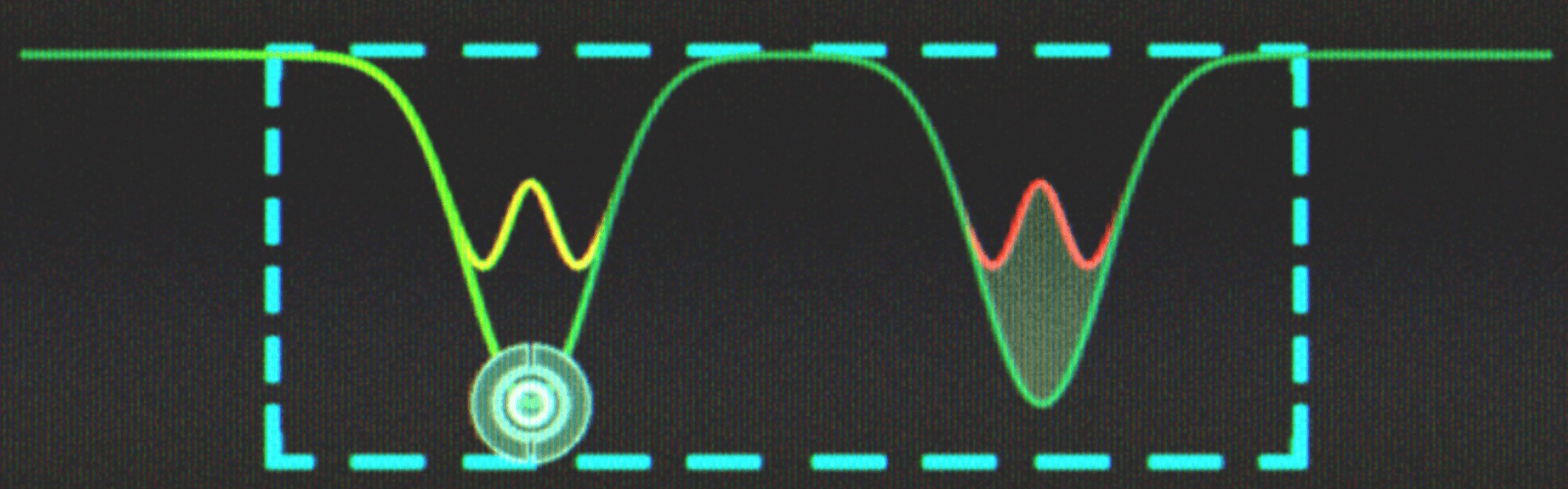}\\
		\vspace{0.25cm}
		\includegraphics[width=\thewidth\linewidth, trim={2.5cm 0cm 2.5cm 7.5cm},clip]{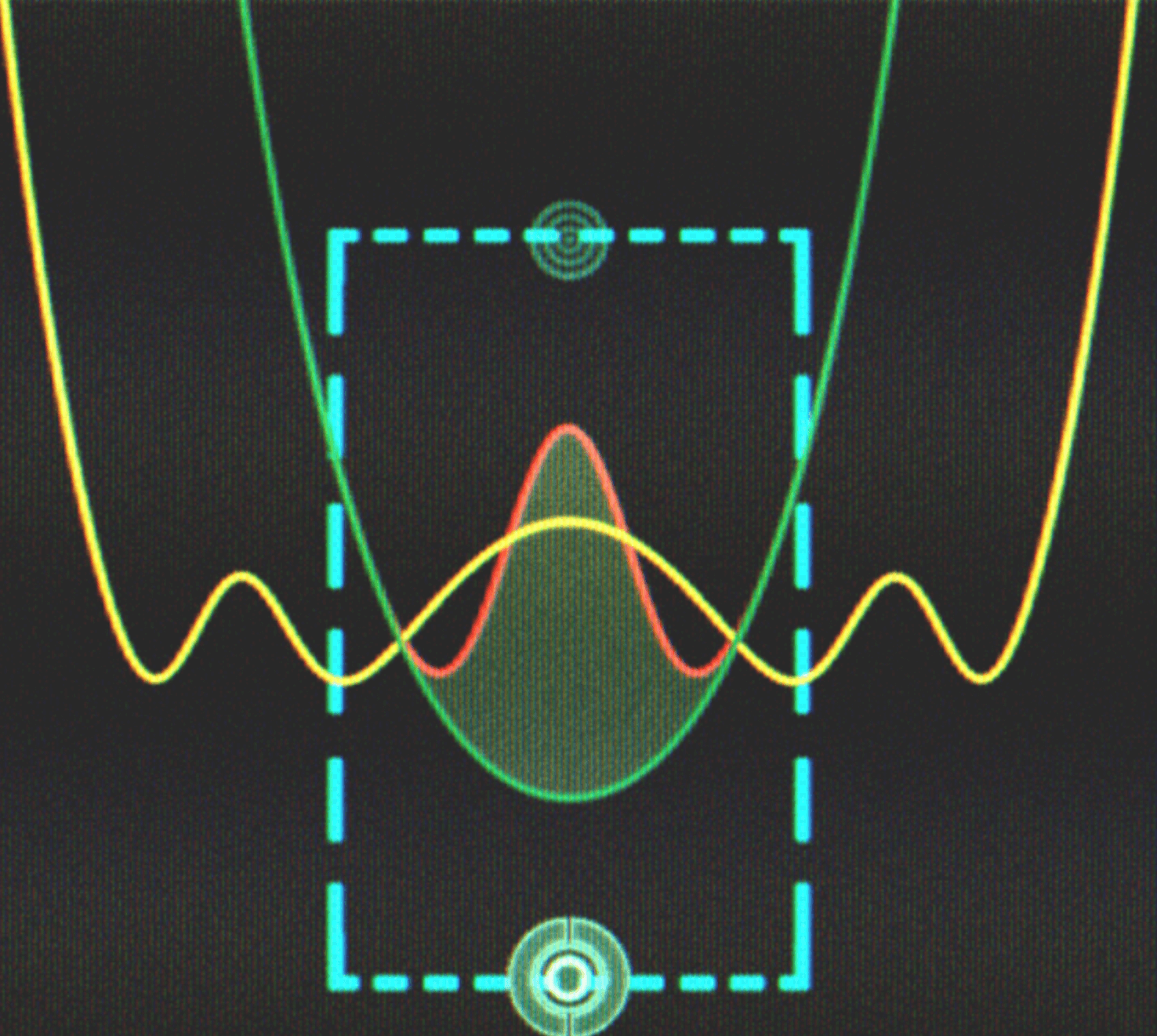}\\
		\vspace{0.25cm}
		\includegraphics[width=\thewidth\linewidth, trim={0cm 9cm 0cm 16.5cm},clip]{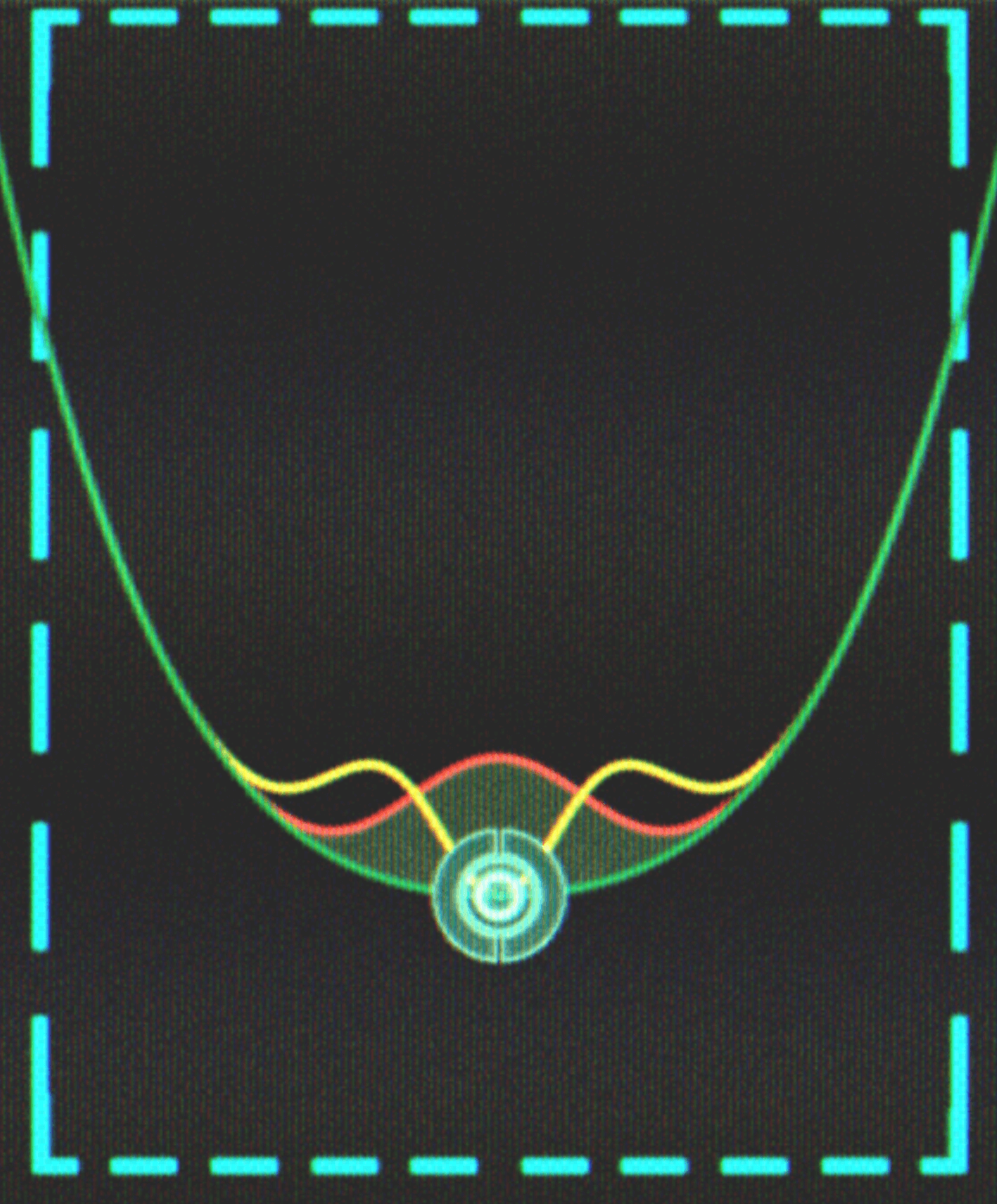}

	\caption{The three central state transfer problems in \textit{Quantum Moves 2}. 
		Top: \textit{Bring Home Water} (single-particle). 
		Middle: \textit{Splitting} (\bec). 
		Bottom: \textit{Shake Up} (\bec).
		The instantaneous density $|\psi(x,t)|^2$ (red line) must be transfered into the target density $|\psitgt (x)|^2$ (yellow line) without residual excitation.
		The trapping potential (green line) is parametrized by the position of the round, draggable cursor, $\left\{u_1(t), u_2(t)\right\} = \left\{f_1(x\st{cursor}(t)), f_2(y\st{cursor}(t)\right\})$, which is unable to leave the turquoise bounding box (control boundaries).
		The functions $f_1$ and $f_2$ are linear. 
		The wave function densities are offset by the potential, $|\psi(x)|^2 + V(x)$,  for illustrative purposes (e.g. the \textit{Splitting} target density represents two equally sized wave packets trapped in a double well -- the large bump in the center is the barrier and two smaller bumps are the wave packets).
	}
	\label{fig:levels}
\end{figure}

Here we briefly describe and motivate the three main levels in the game. Fig.~\ref{fig:levels} displays their in-game representation.
See Appendix~\ref{app:game} for a more complete description of the game interface.
 \\

\pgfmathsetmacro{\theitemindent}{-0.25}
\begin{itemize}[itemindent=\theitemindent em,align=left,  leftmargin=*]
\item[\textit{Bring Home Water}:]
A single atom resides in the ground state of a static tweezer and must be picked up and shuttled back 
into the ground state at the original location of the movable tweezer. This type of transfer is necessary for implementing quantum computations in neutral atoms based on collision gates \cite{weitenberg2011quantum}.

\item[\textit{Splitting}:]
A \bec initially resides in the ground state of a single-well configuration on an atom chip and must be transferred into the ground state of a double-well configuration by deforming the potential. 
The split condensate can then be used for matter-wave interferometry \cite{schumm2005matter,hohenester2007optimal,jager2014optimal,sorensen2018approaching}. 

\item[\textit{Shake Up}:]
A \bec initially resides in the ground state of a single-well configuration on an atom chip and must be transferred into the first excited state by shaking the potential. 
The excited state of the \bec acts as a source for twin-atom beams  \cite{bucker2011twin,van2016optimal,jager2014optimal,sorensen2018quantum,sorensen2018approaching}.
\end{itemize}


\section{Algorithms and Seeding Strategies}
\label{sec:algseed}
In this section we specify the suite of different algorithms and seeding strategies under consideration. 
We define a \textit{method} as a particular combination of algorithm and seeding strategy. For example, \grape \pr-\rs
is the method that uses the \grape algorithm to optimize preselected random seeds. 
Further details of numerics and each algorithm are included in Appendices~\ref{app:numerics}-\ref{app:algorithms}. 

\subsection{Algorithms}

\pgfmathsetmacro{\theitemindent}{-0.25}
\begin{itemize}[itemindent=\theitemindent em,align=left,leftmargin=*]
	\item [\textsl{\grape}:] Standard gradient-based optimization using the \textsc{l-bfgs} quasi-Newton search direction with line search. 
	Bandwidth limitation (smoothness) is included through a derivative-regularization cost term. 

	\item[\textsl{\pgrape}:] Player \grape. The player can start and stop the optimization. Otherwise, it is identical to \grape. The algorithm 
	is executed locally on the player's device as a part of the game.

 	\item[\textsl{Stochastic Ascent (\sa)}:] Gradient-free maximally greedy time-local search. 
 	The process duration $T$ is segmented into $n_b$ bins of equal width within which the control values are constant. The bin values are updated in a stochastic order.
 	The bandwidth limitation (smoothness) is inversely proportional to $n_b$ (there is no additional regularization cost term in the current implementation).
	
\end{itemize}

\vspace{-0.2cm}
\subsection{Seeding Strategies}
\vspace{-0.2cm}
As outlined in Sec.~\ref{sec:qoc}, seeding strategies are a fundamental component of optimization. 
Drawing from random distributions provides the most generic way of seeding \footnote{
	Another common seeding strategy is random seeding in frequency space which has the advantage of enabling frequency cutoffs and thereby adjustable levels of smoothness of the seeds. 
	Although we have also conducted the analysis below for this type of seed parametrization, it will not be presented here in order to preserve the clarity of the presentation, as it generally performed on par with or slightly worse than the uniform random seeding for the chosen parameters.}: 
\begin{itemize}[itemindent=\theitemindent em,align=left,leftmargin=*]
	\item[\textsl{Random Seed (\rs)}:] 
	The control is assembled by independently sampling a uniform distribution within given boundaries (Appendix~\ref{app:numerics}) for each control parameter,
	\begin{align}
		u_p(t) = \mathrm{uniform}(u\st{min},u\st{max}),
	\end{align}
	for $p=1,2$. Subsequent control values at $u_p(t+\delta t)$ are completely uncorrelated. 
	To introduce correlations, one can also segment $T$ into $n_b$ bins of equal width $w$ such that the control value is initially constant within each bin (see also Appendix \ref{app:algorithms}). 
%
\end{itemize}

\noindent Quantum Moves 2 provides two novel seeding strategies:
\begin{itemize}[itemindent=\theitemindent em,align=left,leftmargin=*]		
	\item[\textsl{Player Seed (\ps)}:]
	The control is assembled by mapping the players' cursor position during gameplay as a function of time,
	 \begin{align}
	 	\begin{pmatrix}
	 	u_1(t)\\
		u_2(t)
	 \end{pmatrix}
	 =
	\begin{pmatrix}
	 f_1(x\st{cursor}(t)) \\
	 f_2(y\st{cursor}(t))
	 \end{pmatrix}. \label{eq:f1f2}
	 \end{align}

	\item [\textsl{Player Optimized Seed (\po)}:] This seeding strategy describes a \ps seed that has been optimized by the player (i.e. \po $\equiv$ \pgrape \ps),
	 \begin{align}
		\begin{pmatrix}
		u_1(t)\\
		u_2(t)
		\end{pmatrix}
		=
		\pgrape
		\begin{pmatrix}
		f_1(x\st{cursor}(t)) \\
		f_2(y\st{cursor}(t))
		\end{pmatrix}.
		\end{align}
		If the optimization was stopped before convergence, the optimized control can be used as a seed for further optimization.
		If the optimization converged, the seed itself is already a local optimum.
		From a resource perspective, these seeds are very valuable since they come partially or fully pre-optimized at no cost to the research team.
\end{itemize}
A heuristic extension of any seeding strategy is preselection:
\begin{itemize}[itemindent=\theitemindent em,align=left,leftmargin=*]			
	\item [\textsl{Preselection (\pr)}:] 
	A naïve greedy heuristic to choose which seeds should be picked for optimization.
	Given a set of candidate seeds and their associated fidelities, optimize only the $N$ seeds with highest initial fidelity. 
\end{itemize}


\clearpage

\makeatletter\onecolumngrid@push\makeatother
\begin{figure*}[h]
	\begin{minipage}[b]{\textwidth}
		\centering
		

			\includegraphics[]{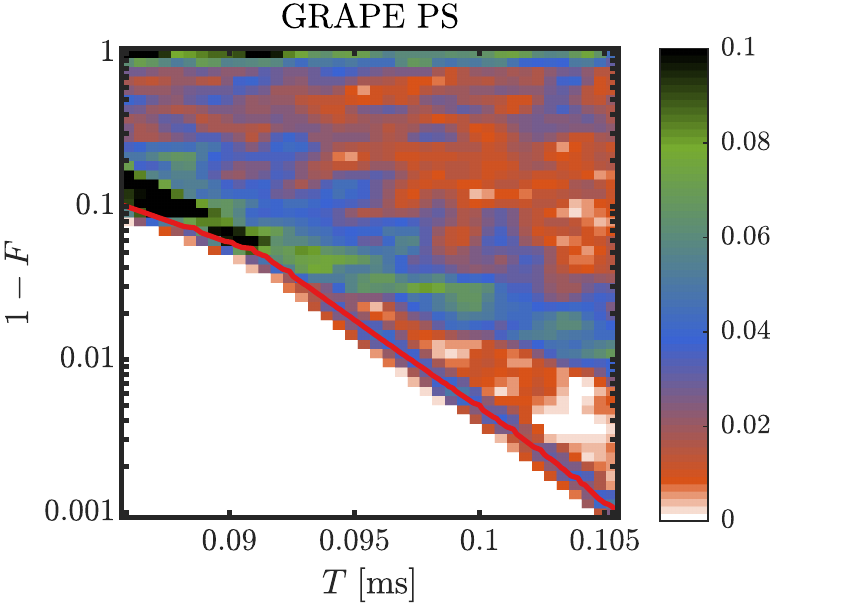}
			\includegraphics[]{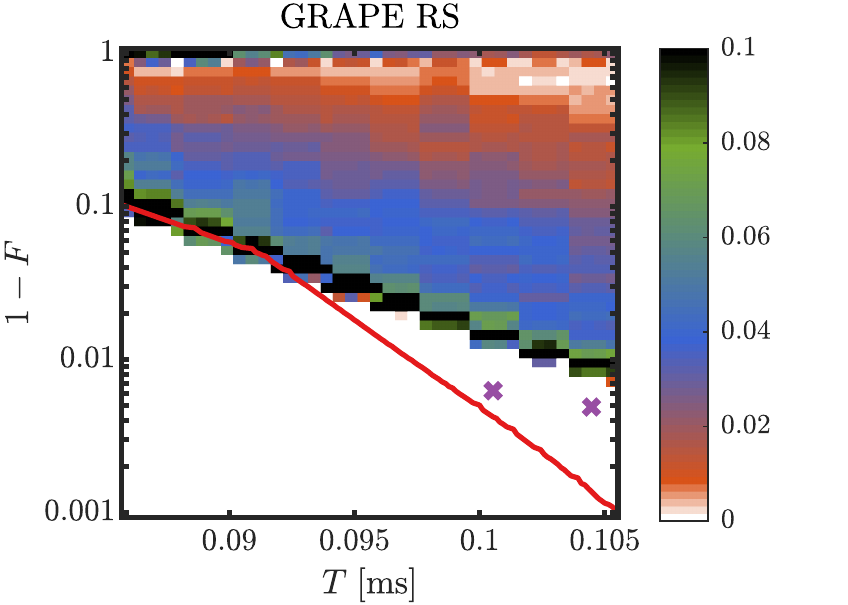} \\
			\includegraphics[]{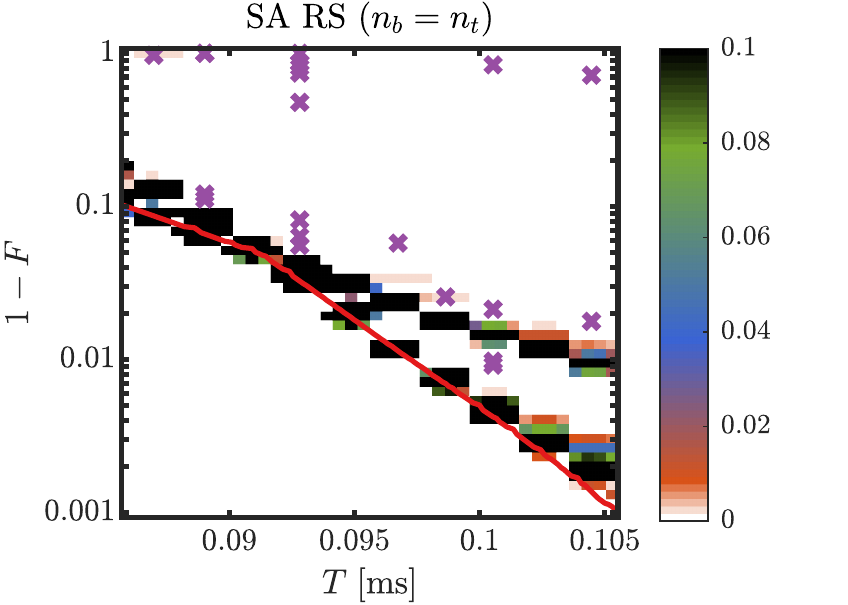}
			\includegraphics[]{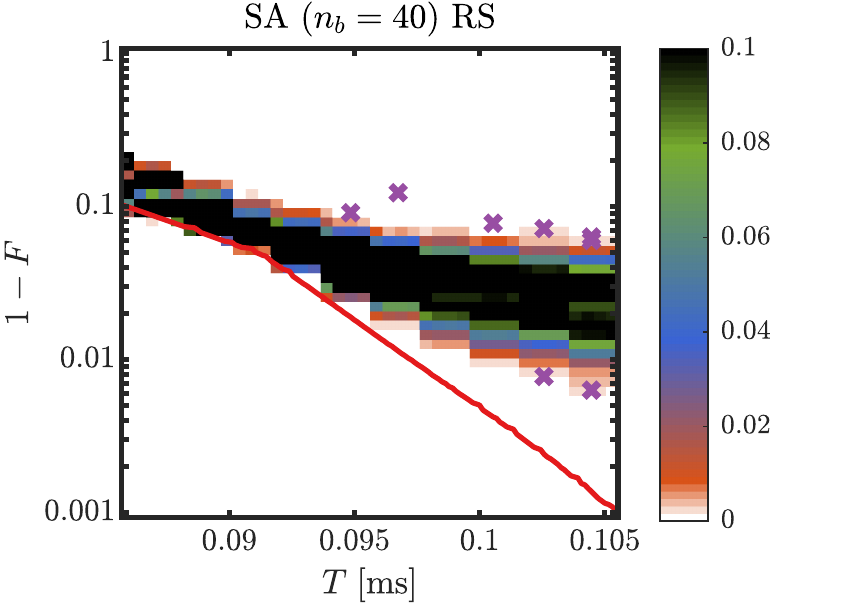} 

		\caption{\textit{Bring Home Water} solution densities for different methods. Each density is normalized for every individual $T$ and thus represents an estimate of the probability distribution for obtaining a particular $F$ for a given $T$ ($\mathcal{P}(F|T)$). The reference curve shows the best obtained results for \grape \ps. Purple crosses indicate individual solutions at densities lower than 0.002. 
		}
		\label{fig:solutiondensities_bhw}
		\hspace{-1.75cm}
				\includegraphics[]{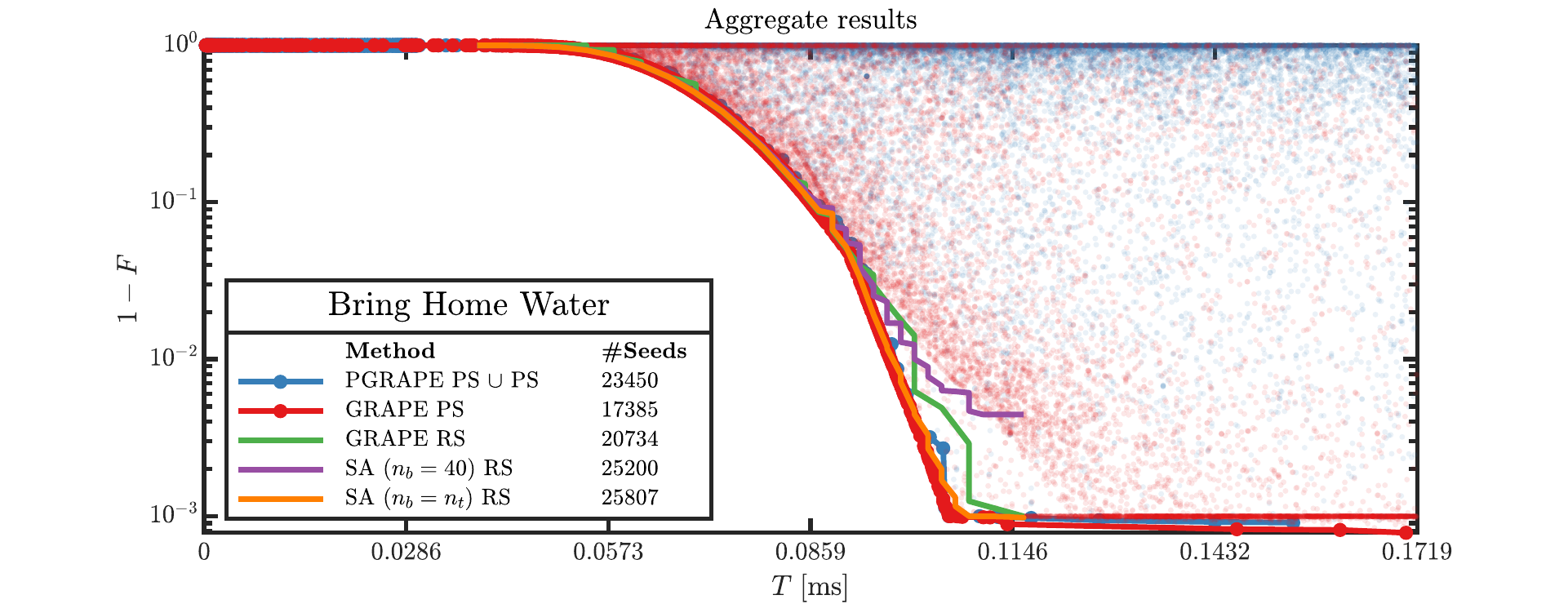}

		\caption{Aggregate, monotonically best optimization results (lower is better) for several methods in \textit{Bring Home Water}. 
			Solid lines with (without) dots show the best results obtained with (without) player influence.
			The scattered blue dots show all results produced by 536 players seeding on average $\sim 32$ solutions and optimizing approximately $1/4$ of these for $\sim 131$ iterations on average.  
			The scattered red dots show the same, except the optimization is carried out on all player seeds with the computational resources described in the text (i.e. no player influence after seeding). The dot translucency indicates the density distribution. 
		}
		\label{fig:FT_bhw}

	\end{minipage}
\end{figure*}
\clearpage
\makeatletter\onecolumngrid@pop\makeatother

\section{Bring Home Water}
\label{sec:bhw}
In this section we present optimization results obtained in \textit{Bring Home Water} for the different methods described in Sec.~\ref{sec:algseed}. The alloted resources for the optimizations are discussed in Appendix~\ref{app:algorithms}.
We denote by \pgrape \ps $\cup$ \ps the joint set of optimized and unoptimized solutions produced solely by the players in-game
\footnote{
	Due to technical limitations at the time, all optimizations of player solutions (inside and outside the game) in \textit{Bring Home Water} were performed with the 
	same numerical implementation of Eq.~\eqref{eq:H} as $g\neq 0$ but with $g=0$. This (artificially) increases the operational costs compared to optimization of \rs solutions. As we shall see in the following, however, the conclusions are robust towards this slight skew and fully leveling the playing field would serve to reinforce them. 
}.


Figure~\ref{fig:solutiondensities_bhw} shows 
the density distribution of solution infidelities $(1-F)$ 
\footnote{The densities are obtained by kernel density estimation using the Epanechnikov (parabolic) kernel with bandwidth 0.08 in the infidelity dimension (evaluated as $\log_{10}(1 - F)$), which attributes a localized density peak to each sampled data point and yields a density map when adding all the points. } 
as a function of process duration $T$ in the high-fidelity regime for various methods.
For a given method, the solution densities estimate the probability distribution $\mathcal{P}(F|T)$ of obtaining a particular fidelity for a given $T$, since each column is individually normalized.
For reference, the solid line shows the best obtained \grape \ps results from Fig.~\ref{fig:FT_bhw}. 
Figure~\ref{fig:FT_bhw} shows the aggregate, monotonically best infidelity as a function of process duration $T$. 
The blue dots show all results produced only by players in-game, i.e. player seeds (\ps) and player-optimized seeds (\pgrape \ps). 
The red dots show the result of \grape \ps, i.e. computer cluster optimized player seeds (\ps) only (inducing a difference in number of seeds in Fig.~\ref{fig:FT_bhw}).
Lines with (without) dots show the best results obtained by methods based on player (random) seeds.

For the \grape \ps density and \pgrape \ps $\cup$ \ps results shown in Fig.~\ref{fig:FT_bhw}, we observe two clear bands of solutions that are each described by distinct exponential behavior, which hints at two corresponding solution strategies.
This is verified and analyzed in Sec.~\ref{sec:clustering_bhw} using clustering techniques. 
The identification of the two exponentially-gapped solution strategies and the relative likelihoods of different methods identifying each strategy is the first of three main findings in this section.
The duration-dependent globally optimal strategy changes from one to the other near $T = 0.092\,\si{ms}$, explaining the kink in the best result reference curve seen in Fig.~\ref{fig:solutiondensities_bhw} and the departing line of red dots in Fig.~\ref{fig:FT_bhw}.
The gap between the global and local optimal strategies increases exponentially, leading to significantly different quantum speed limit $T\st{QSL}^{F}$ estimates (as defined in Sec.~\ref{sec:qoc}). 
Outside of these strategies, the densities are mostly sparse but non-zero, indicating a topographically complex optimization landscape containing many isolated local traps or regions with near-vanishing gradient. 
This is also understood by the interspersed red and blue dot distributions in Fig.~\ref{fig:FT_bhw}.

We now examine \grape \rs, where the only difference with respect to the former methods is the seeding strategy.
In this case, we observe that only the inferior, locally optimal strategy is discovered, with only two (out of 20734) solutions 
located in the strategy gap. Evidently, for the same optimization algorithm, the structure of player seeds is preferable. The deficiency of \grape \rs is analyzed in Sec.~\ref{sec:grapers_correlations}.

Next, we turn our attention to \sa \rs variants with full resolution $(n_b = n_t)$ and reduced resolution $(n_b=40)$.
As discussed in detail in Appendix~\ref{app:algorithms}, \sa is expected to be efficient for linear problems ($g=0$) with preferably a single control parameter.
For \textit{Bring Home Water} the former is satisfied by the problem definition and the latter can be satisfied by choosing the tweezer amplitude control such that it is maximally deep at all times, $u_2(t) = u_2\ut{min}$, as is also done in Ref.~\cite{sels2018stochastic} and \sa thus only optimizes $u_1(t)$  
\footnote{
	This \textit{a priori} fixation heuristic is the strongest type of structure to impose on the problem, the motivation of which is rooted in insight regarding the particular problem (steeper potentials yield faster dynamical time scales). 
	This type of heuristic may not always be readily available, e.g. if the control parameters must simultaneously change in a non-trivial way to facilitate the transfer (e.g. the $\theta(t)$ and $\beta(t)$ parameters in Ref.~\cite{jensen2019time}). 
	Nevertheless, 
	here we choose the \textit{a priori} heuristically constrained version to be consistent with previous studies \cite{sels2018stochastic,gronlund2019algorithms}. Likewise, the number of bins for the reduced resolution variant, $(n_b = 40)$, was chosen based on these previous studies. 
}.
Generally, in all the optimized \grape solutions, we indeed observe that the tweezer attains its maximal depth whenever the tweezer is in contact with the atom (overlap with non-zero probability density).

With these choices, \sa \rs ($n_b = n_t$) is capable of discovering the two distinct strategies with remarkable efficiency, despite the structureless nature of the $u_1(t)$ seeding mechanism. Only a few low fidelity solutions occur.
\sa \rs ($n_b = 40$) does not identify the globally optimal strategy (lower branch) and instead concentrates on a broad band of solutions with a center shifted to above the inferior strategy (upper branch).
This can be attributed to the reduced resolution of the algorithm preventing it from resolving the dynamics finely enough 
\footnote{
It is safe to conjecture that a more careful choice of the heuristic bin parameter $40 < n_b < n_t$ would yield results matching the full resolution ($n_b = n_t$) above some resolution threshold.
}.

Both \grape \ps and \sa \rs ($n_b=n_t$) find the same estimates $T\st{QSL}^{F=0.99} \approx 0.0973\,\si{ms}$ and $T\st{QSL}^{F=0.999} \approx 0.1057\,\si{ms}$, whereas the estimate from \pgrape \ps $\cup$ \ps is off by less than $1\%$ with respect to $F=0.99$. 
The fact that the in-game player-optimized curve roughly matches the best results optimized on the computer cluster represents the second main finding of this section. This, coupled with similar findings for the two remaining challenges, represents our quantitative confirmation of Q1 (cf. Sec.~\ref{sec:introduction}) for this range of control problems.

\grape \ps,  \pgrape \ps $\cup$ \ps, and \sa \rs ($n_b=n_t$) are also all able to find both solution strategies, but  \sa \rs ($n_b=n_t$) is the most efficient at optimizing low-fidelity solutions into high-fidelity solutions. This cannot be attributed to the different seeding mechanisms, since \grape \rs fails to find the globally optimal solution strategy. 
Instead, the difference is due to how the two optimization algorithms traverse the landscape (Appendix~\ref{app:algorithms}) and how they 
respond to being near a local trap 
\footnote
{
\sa performs exhaustive search (within its discretization) along a single, stochastically chosen dimension at a time and
it is therefore possible to escape from initial trapping regions likely to be found by structureless random seeding. 
Eventually, the optimal strategies are discoverable given that their attractors are sufficiently broad (which for an arbitrary problem is not generally guaranteed). 
On the other hand, \grape is principally a strict local optimizer and will rapidly converge to the local trap. 
Thus, when using \grape the structureless \rs seeding is more likely to get stuck in local traps, whereas the more structured \ps seeding is more likely to yield better results. 
}.

Later in Sec.~\ref{sec:sampling} we provide an alternative statistical characterization of each method, as well as the effect of the preselection heuristic, for each of the three problems.


\subsection{Optimal Strategies -- Control Clustering}
\label{sec:clustering_bhw}


In order to extract the identified solution strategies, we apply \dbscan clustering \cite{ester1996density} to the \grape \ps method. Based on the results presented in the previous section, 
we expect the existence of distinct solution strategies, i.e. families of solutions that have a similar functional shape and characteristics but possibly different durations 
\footnote{
\label{footnote:bhwcluster}
The \dbscan clustering algorithm allows controls that are far away from each other (as measured by, e.g., the Euclidean metric) to be in the same cluster, provided there are other controls bridging the gap between them.
This property enables solutions from a given optimal strategy to be grouped within the same cluster.
}.

The clustering was performed only on the duration-normalized tweezer position, $u_1(t/T)$, justified by the fact that the optimized $u_2(t)$ control is, in general, maximally deep. To simplify the clustering, all solutions were given the same number of points by linearly interpolating on a 1000 point grid (the original number of points was defined by $T/\delta t$). For this analysis, we selected high-fidelity solutions $F \in [0.95, 0.999]$ with $T\in [0.093,  0.124 ]\,\si{ms}$ (i.e. near $T\st{QSL}^{F=0.99}$ and $T\st{QSL}^{F=0.999}$).
We used the \verb|sklearn| implementation of \dbscan with Euclidean metric, $\epsilon = 3$, and a minimum number of neighbors $\mathrm{min}\st{samples} = 5$. 

\begin{figure}[t]
	\centering

				\includegraphics[]{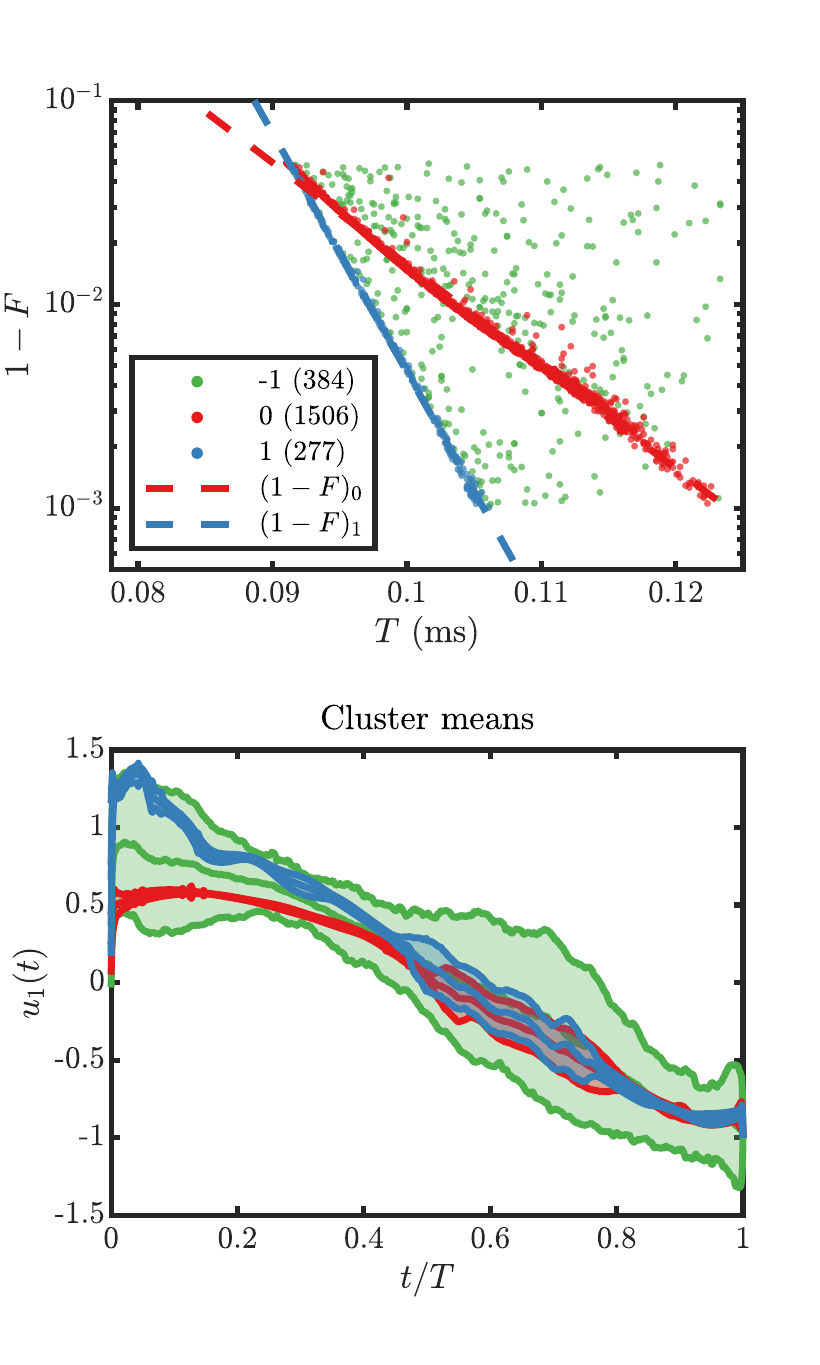}	

	\caption{\textit{Bring Home Water}: Clustering of controls. 
		The cluster color code and their population are shown in the legend.
		Top: Each cluster (0 and 1)	corresponds to a strategy and the gap between them exhibits an exponential $1-F(T)$ behavior, leading to different estimates of quantum speed limits for a given threshold. The unclassified points (-1) are mostly populated by controls similar to either strategy, except for a few local defects that makes them appear distant to the cluster with respect to the Euclidean metric.		
	Bottom: Lines correspond to cluster means and shaded areas to the standard deviation. The most populous cluster (0) corresponds to a front-swing strategy with the tweezer immediately being placed in front of the atom whereas the less populous cluster (1) corresponds to a back-swing strategy with the tweezer immediately being placed behind the atom. 
}
	\label{fig:clusterremovedelay}
	\vspace{0cm}
\end{figure}

After filtering out physically insignificant delays 
\footnote{
	Initial results revealed two major and three minor clusters. By inspection it was found that each smaller cluster was in fact identical to one of the major clusters, apart from an initial delay in placing the tweezer near the atom.
	In order to filter out this physically insignificant delay we apply a correction procedure to all the controls, dropping all initial points before the first instance of $u_1(t/T) \geq 0$. 
	Applying the clustering with the same clustering parameters returns a significantly simpler result as seen in Fig.~\ref{fig:clusterremovedelay}. 
	Examination of the unclassified controls suggest that those solutions do belong to one of the two strategies but have a few local defects (sudden displacements) that make them sub-optimal with respect to the nearest cluster and appear far away from other cluster members based on the Euclidean metric. Thus, effectively only two (high-fidelity) optimal strategies exist. 
} 
we find two major clusters (labeled 0 and 1) with increased populations and a ``cluster'' (labeled -1) in which local defects cause irreparable deviations from the two strategies.
Each cluster exhibits a strikingly exponential trade-off between fidelity and duration well-described by the fits
\begin{align}
(1-F(T))_0 &= 10^{-1.45 - 50.11\cdot (T/\si{ms} - 0.0929)},
\\
(1-F(T))_1 &= 10^{-1.50 - 117.27\cdot(T/\si{ms} - 0.0929)}, 
\end{align}
and it is clearly seen that the strategy gap widens exponentially. 
Assuming these trends can be extrapolated, this yields $T\st{QSL,0}^F / T\st{QSL,1}^F = 1.17, 1.26, 1.33$ 
and infidelity ratios of $1\cdot 10^2,\,3\cdot 10^3,\, 6\cdot 10^4$ at $T\st{QSL,0}^F$ for $F=0.999, 0.9999, 0.99999$, respectively. 



Physically, the strategy corresponding to the cluster 0 begins by placing the tweezer in front of the atom, providing immediate acceleration towards the target position. We name this the \textit{front-swing} strategy.
Conversely, the strategy corresponding to the cluster 1 begins by placing the tweezer behind the atom, providing immediate acceleration away from the target position. We name this the \textit{back-swing} strategy.
In both instances, there is very little deviation from the cluster mean except during the shuttling of the atom where small deviations are allowed.

Intuitively one might expect that the back-swing strategy would be slower because the atom must travel an overall longer distance compared to the front-swing strategy, but this is evidently not the case. 
Instead, initially displacing the atom onto the static well's right side can serve as an additional accelerating force to that of the movable tweezer.
With this analysis, we are in a position to explore in the following section why \grape \rs fails to locate the back-swing strategy. 


\subsection{GRAPE RS -- Efficiency vs Optimality}
\label{sec:grapers_correlations}
It is clear from Fig.~\ref{fig:FT_bhw} that the fine-grained \grape \rs (with $n_b = n_t$) fails to find the back-swing solution 
where the same optimization algorithm with \ps succeeds. 
As shown in Sec.~\ref{sec:clustering_bhw}, this strategy requires the control to be immediately placed on the right-hand side of the static tweezer $(u_1 > x_0)$ for a time interval sufficient to displace the atom.
To gain a sense of timescales, the harmonic approximation to the fully overlapped tweezers $(u_1 = x_0)$ yields an oscillation period of $T =2\pi/\omega \approx 0.1$ (in simulation units, see Appendix \ref{app:numerics}). 
Assuming $T/10$ is a sufficient response time for meaningful dynamics, the probability for consecutively randomly sampling the corresponding number of points $n = (T/10)/\delta t =  10^{-2}/(3.5\cdot10^{-4}) \approx 29$ such that they all have $(u_1 > x_0)$ is vanishingly small $P(u_1 > x_0)^n =(0.25)^n = 6\cdot 10^{-18}$ when successive control values are uncorrelated. Even at a much less conservative estimate of $T/40$, the probability $5 \cdot 10^{-5}$ remains strongly suppressed. 
Since \grape traverses the landscape locally, it is therefore almost guaranteed that subsequent optimization leads to the front-swing strategy. The same is not observed for \sa, since the one-dimensional exhaustive search allows for the individual adjustment of points $(u_1 < x_0)$ into $(u_1 > x_0)$. 

\begin{figure}
	
			\includegraphics[]{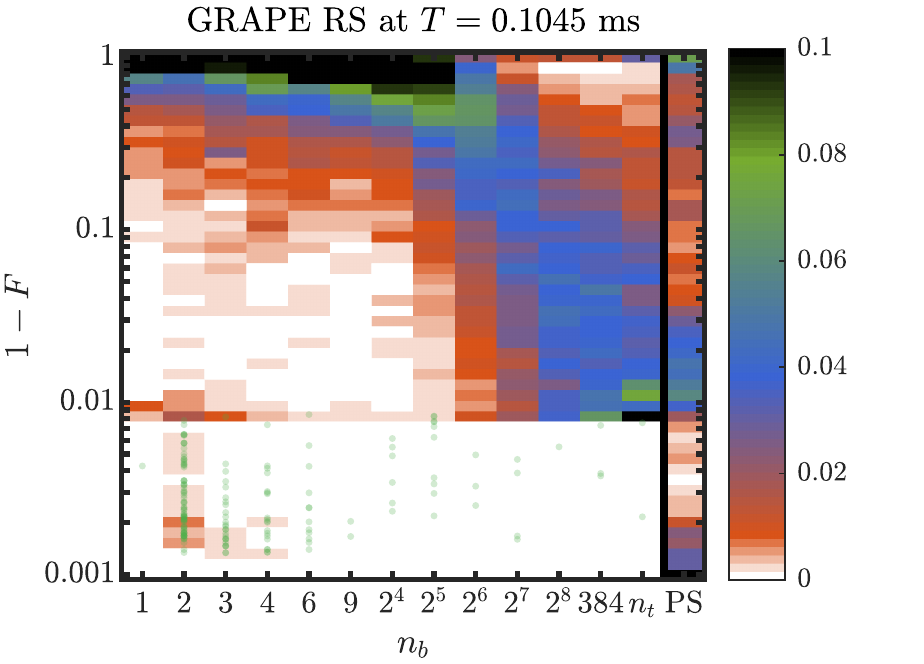}
			\includegraphics[]{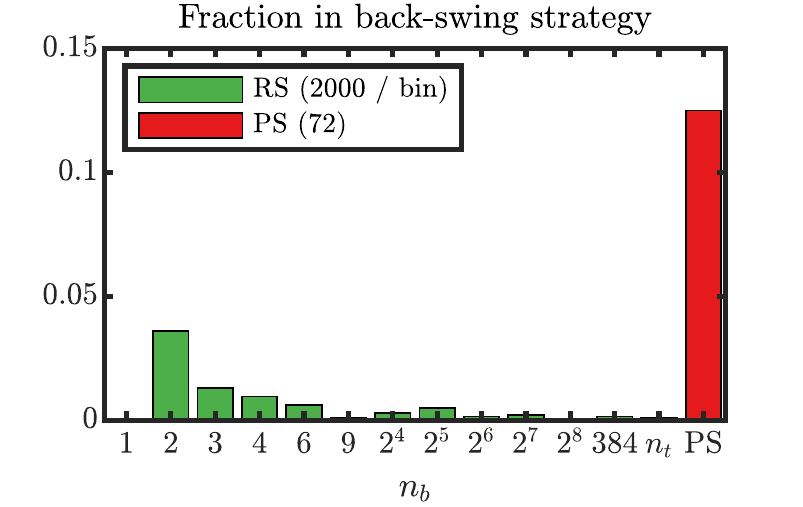}
	\vspace{-0.3cm}
	\caption{Top: Solution densities as a function of number of bins $n_b$ at $T = 0.1045$.
		Each green dot denotes a solution in the back-swing strategy.
		The density in the right black box shows the \grape \ps solution density from Fig.~\ref{fig:solutiondensities_bhw} at the same duration denoted by \ps on the $x$-axis (the duration is on the edge of one of the bin limits and the density is therefore taken as the mean of the two neighboring bins). 
		Bottom: Histogram of the number of back-swing solutions, counted as those with fidelity higher than the best front-swing result at $n_b = n_t$, for \grape \rs with variable $n_b$ (green) and \grape \ps (red) from Fig.~\ref{fig:FT_bhw}. The total number of seeds per bin is indicated in the parentheses. 
	}
	\label{fig:grapers_correlations}
	\vspace{-0.3cm}
\end{figure}

One possible explanation for the observed difference between randomly and player-seeded behavior is thus that with a very high time resolution the random seeds oscillate rapidly whereas the player seeds are typically comparatively rather smooth due to the physical limitations in the speed of the players' cursor movements.
One might then hypothesize that sufficiently coarse-grained piecewise constant random seeds 
would also yield similarly good behavior. If this was the case, the player superiority would indeed be a trivial artifact of the choice of discretization. To test this, we therefore
study the behavior of \grape \rs optimizations when heuristically introducing correlations in the random seed by dividing the seed into $n_b\leq n_t$ piecewise constant control segments (like with \sa, but in this case the subsequent optimization is not constrained to this parametrization). This makes it increasingly likely for the seed to initially begin and remain on the right side of the atom for a sufficient amount of time. Figure~\ref{fig:grapers_correlations} shows the results of optimizing 2000 of these seeds near $T\st{QSL}^{F=0.999}$ as a function of $n_b$, as well as the \grape \ps results from Figs.~\ref{fig:solutiondensities_bhw}-\ref{fig:FT_bhw}.
In the limit $n_b = n_t$, there is a high probability of finding solutions belonging to the front-swing strategy with associated fidelities around $0.99$. This is consistent with the findings in Fig.~\ref{fig:solutiondensities_bhw}.
Only for $n_b \lesssim 4$ is there an appreciable albeit low probability (about 1 to 3.5\%) of identifying back-swing solutions with much higher associated fidelities above $0.99$, whereas more than $91\%$ of the density resides below $F=0.9$. 
The increase in low fidelity solutions can be attributed to the fact for e.g. $n_b=2$ (that has the highest empiric probability of finding the back-swing), there is an increased probability of placing the control tweezer far away from the atom and never touching the atom at all, leaving it in the initial state and resulting in a vanishing gradient (the gradient is proportional to $\braket{\psitgt|\psi(T)}$). 
On the other hand, the probability of \grape \ps finding the back-swing strategy is much larger (about 13\%) with only $36\%$ of the solutions below $F=0.90$. 
The non-trivial observation that player seeds, for the same optimizer, outperform randomly generated piecewise constant seeds at all coarseness scales represents the third main contribution of this section. 

The shapes of both optimal strategies in Fig.~\ref{fig:clusterremovedelay} contain a major linear component, and this suggests prospects for improvement by employing piecewise linear rather than piecewise constant seeding. 
That is, a figure similar to Fig.~\ref{fig:grapers_correlations} with results from \grape employing such seeds would results would therefore likely yield a dramatically different picture. 
In particular, we note that the exclusive identification of the inferior front-swing strategy from the presented \grape \rs results by itself points to the piecewise linear seeding, which, if applied in subsequent optimization, would likely lead to the discovery of the globally optimal back-swing strategy. 
This relies on the fact that both solution strategies are captured within the parametrization of few piecewise linear segments. 
In general, such a similarity between distinct solution strategies is not guaranteed.
As an example, we find in Sec.~\ref{sec:shakeup} that the \textit{Shake Up} problem also contains a plurality of strategies that are related to one another in a more subtle way. Bulk analysis of solutions from one strategy would thus not necessarily yield a problem parametrization that also encapsulates the other strategies.
This underscores the potential usefulness of having data sets available that have been generated with 
sets of basic underlying assumptions that are as different as possible.

Our conclusion is therefore not that a standard, heuristic-free numerical approach \textit{will} fail on the problem but that it \textit{can} fail, and this
result should not be taken as proof that players can guarantee computational

\clearpage
\makeatletter\onecolumngrid@push\makeatother
\begin{figure*}[t]
	\begin{minipage}{\textwidth}
		
		\centering
				\includegraphics[]{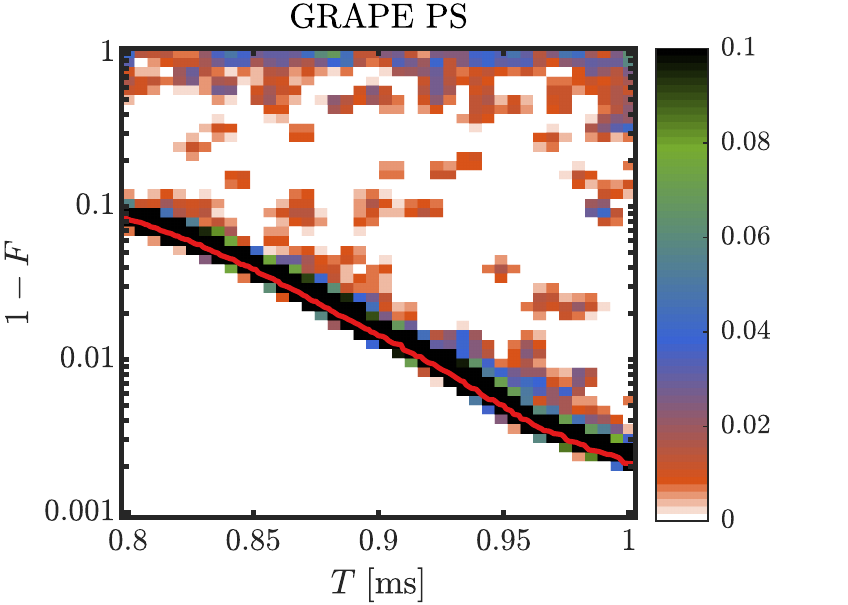}
				\includegraphics[]{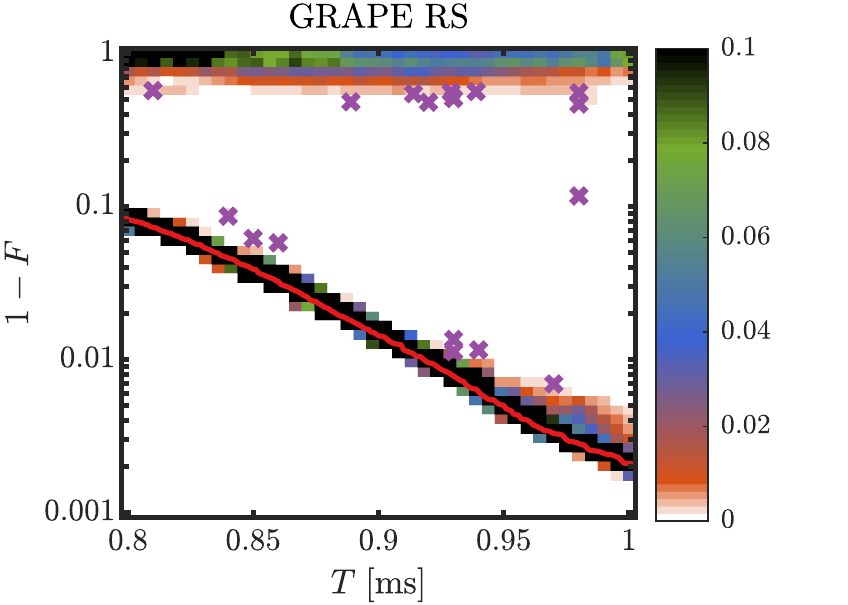} \\
				\includegraphics[]{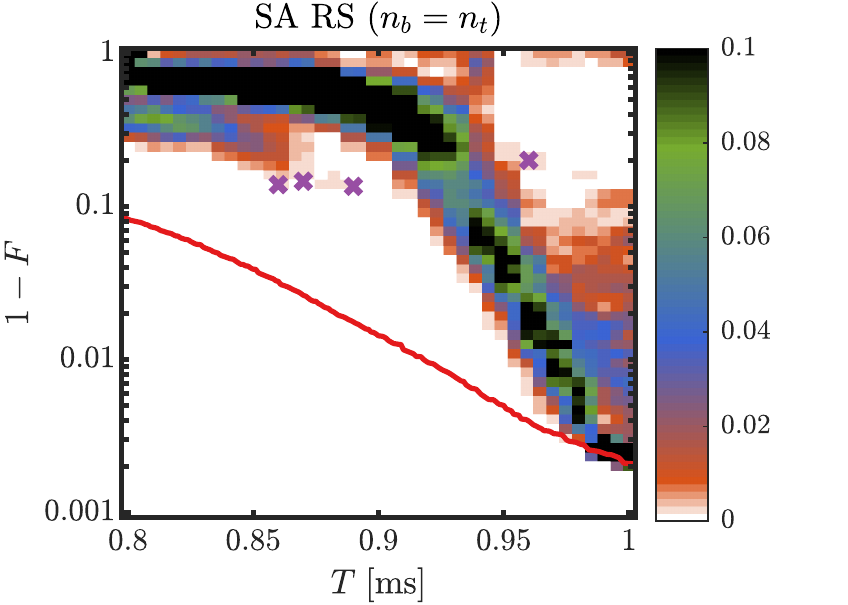}
				\includegraphics[]{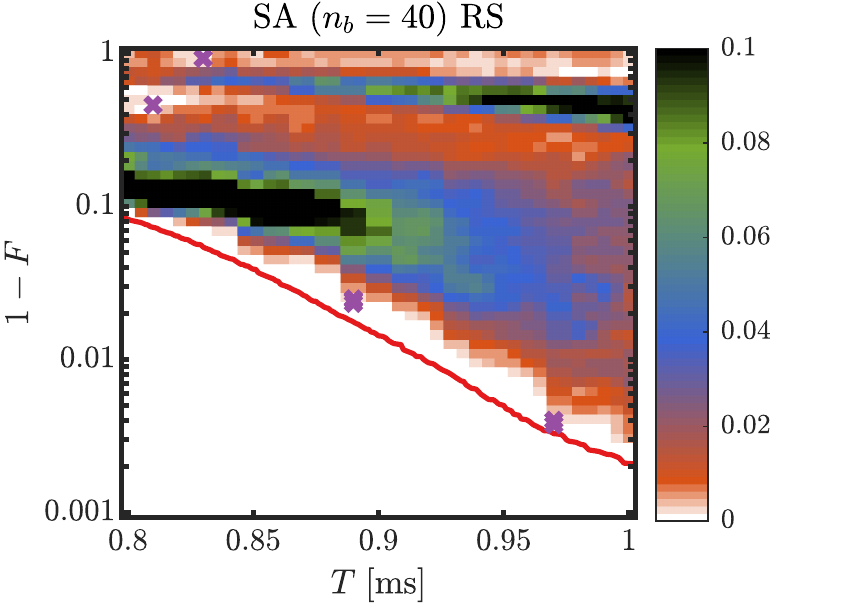} 
		
		\caption{\textit{Splitting} solution densities for different methods. Each density is normalized for every individual $T$ and thus represents an estimate of the probability distribution for obtaining a particular $F$ for a given $T$ ($\mathcal{P}(F|T)$). The red reference curve shows the best obtained results for \grape \ps. Purple crosses indicate individual solutions at densities lower than 0.002.
		}
		\label{fig:solutiondensities_splitting}
		\hspace{-1.75cm}
				\includegraphics[]{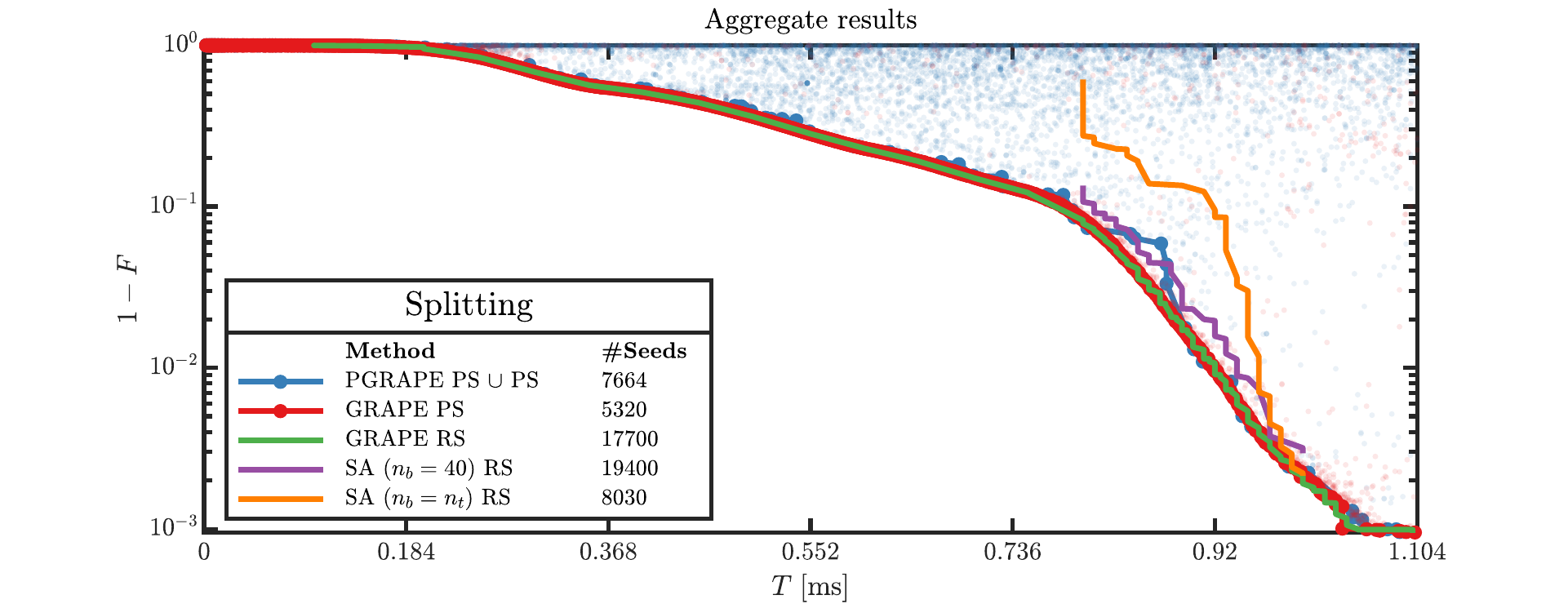}

		\caption{Aggregate, monotonically best optimization results (lower is better) for several methods in \textit{Splitting}. 
			Solid lines with (without) dots show the best results obtained with (without) player influence.
			The scattered blue dots show all results produced by 193 players seeding on average $\sim 28$ solutions and optimizing approximately $1/3$ of these for $\sim 127$ iterations on average.  
			The scattered red dots show  the same, except the optimization is carried out on all player seeds with the computational resources described in the text (i.e. no player influence after seeding). The dot translucency indicates the density distribution. 
		}
		\label{fig:FT_split}
		
	\end{minipage}
\end{figure*}
\clearpage
\makeatletter\onecolumngrid@pop\makeatother

\noindent improvements to this or other problems. It does, however, constitute a necessary first demonstration of the value of examining the gamified approach further and comparing to or integrating it with more sophisticated expert heuristics.

\section{Splitting}
In this section, we present optimization results obtained for the \textit{Splitting} level using the same suite of methods, convergence criteria, and computational resources as in Sec.~\ref{sec:bhw}. 

The solution densities are shown in Fig.~\ref{fig:solutiondensities_splitting} and the aggregate results are shown in Fig.~\ref{fig:FT_split}. 
Looking at the solution density for \grape \ps we see a single, dominant band of solutions with only a sparse population of solutions away from it. 
This hints at a simple, almost trap free, easily-navigated control landscape containing a very broad attractor for a single optimal strategy. 
Indeed, this is also understood by observing the dot density in Fig.~\ref{fig:FT_split}: almost all the red dots coincide with the best curve.
A similar situation can be observed from \grape \rs, except for a modest increase in low fidelity solutions but with virtually no population between these and the optimal band. 
Specifically, the average total density per bin below
{ $F \approx 0.9$} is $(18 \pm 4)\%$ for $\rs$ and $(11 \pm 5)\%$ for \ps.
Player seeds are thus slightly more likely to be within reach of the optimal attractor.
Both methods obtain estimates $T\st{QSL}^{F=0.99} \approx 0.92\,\si{ms}$ and $T\st{QSL}^{F=0.999} \approx 0.105\,\si{ms}$.
The estimates from the fully player-generated \pgrape \ps $\cup$ \ps are off by less than $2\%$ in both instances.

The \sa methods perform significantly worse than the \grape methods on this problem because $g\neq 0$.
Even the best \sa ($ n_b = 40$) results do not reach the same fidelity, possibly due to a combination of reduced controllability (low resolution) and the computational penalties associated with $g\neq 0$ as described in Appendix~\ref{app:algorithms}. 
The full resolution \sa ($n_b = n_t$) also fails to converge almost everywhere, except at $T$ sufficiently larger than $T\st{QSL}^{F=0.99}$. 
This reaffirms that the control landscape topography is benign enough that even an inefficient algorithm can find the optimal strategy with appreciable probability at these durations.

\subsection{Optimal Strategies -- Control Clustering}
\label{sec:clustering_splitting}
\begin{figure}

		\includegraphics[]{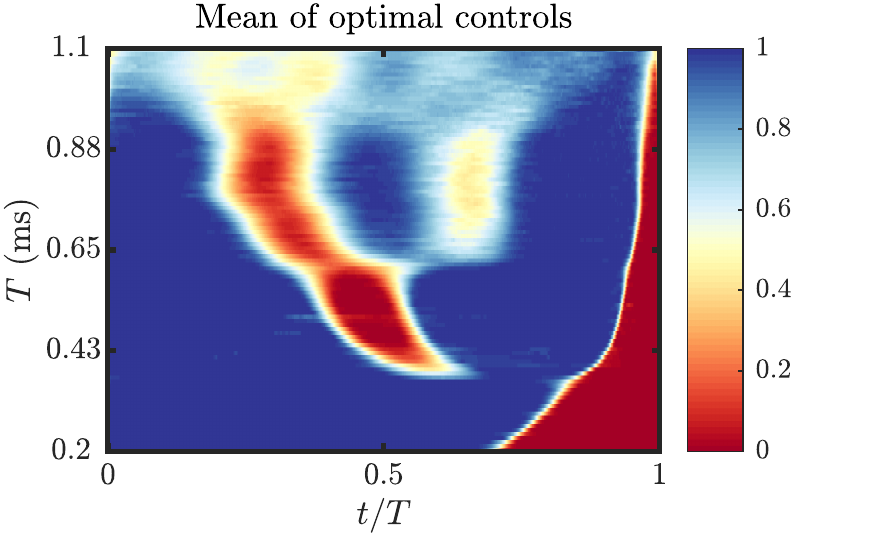}
	\caption{\textit{Splitting}: mean of (near) optimal solutions $\braket{u_2(t/T)}\st{opt}$, corresponding to the height of the potential barrier. For a given $T$ ($y$-axis), 
		the color indicates the mean optimal control value at a given $t/T$ ($x$-axis).
		Clustering analysis identified only a single cluster. 
	}
	\label{fig:splitting_clustering}
\end{figure}

Even without a clustering analysis, the optimal strategy was apparent. 
Figure~\ref{fig:splitting_clustering} shows the mean of all controls within 0.02 of the globally optimal fidelity as a function of $T > 0.2 \,\si{ms}$.
Performing clustering (not shown) with $\epsilon = 5$ and $\mathrm{min}\st{samples} = 5$ on the \grape \ps or \grape \rs results verifies that this problem has a simple optimization landscape: 3569 of the controls were associated with a single cluster, whereas 56 controls were unclassified. 

For all $T$, the mean takes an initially (near) maximal control value for an extended period. This physically corresponds to raising an as steep as possible barrier in the center of the potential (where the wave function is initially localized, see Fig.~\ref{fig:levels}) and thus providing maximal acceleration to split the condensate into two equal wave packets. 
At low $T$ the mean control exhibits a bang-bang structure that tapers off as $T$ increases.
Near $T=0.4\,\si{ms}$ a new bang with smoothed edges appears roughly centered on $t/T\sim 0.5$, which is then bimodally split around $T=0.65 \,\si{ms}$. The mean control then becomes increasingly ``blurry'' indicating a growing departure from bang-bang structures.  


\section{Shake Up}
\label{sec:shakeup}
In this section, we present optimization results obtained in \textit{Shake Up} for the same suite of methods, convergence criteria, and computational resources as in Sec.~\ref{sec:bhw}.

\begin{figure*}[t]
	\begin{minipage}[b]{\textwidth}
		\centering
		

			\includegraphics[]{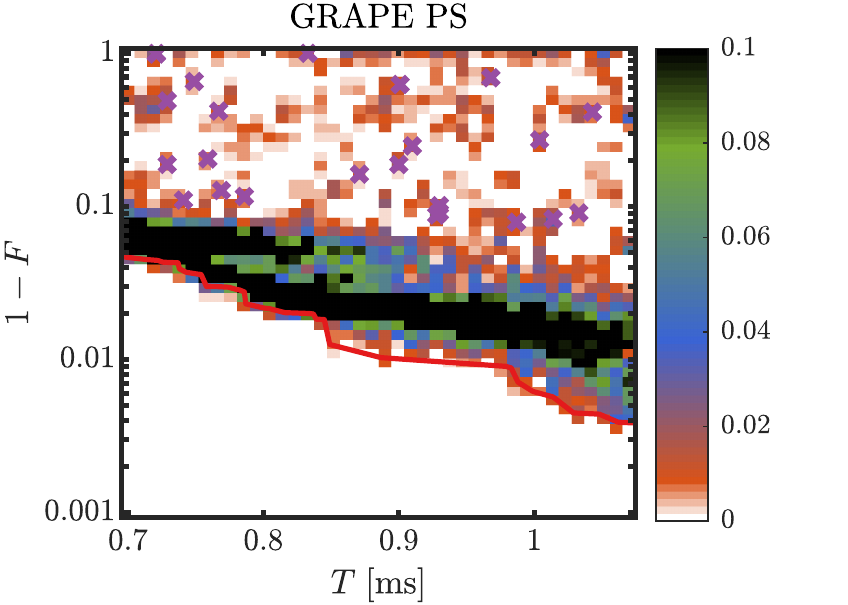}
			\includegraphics[]{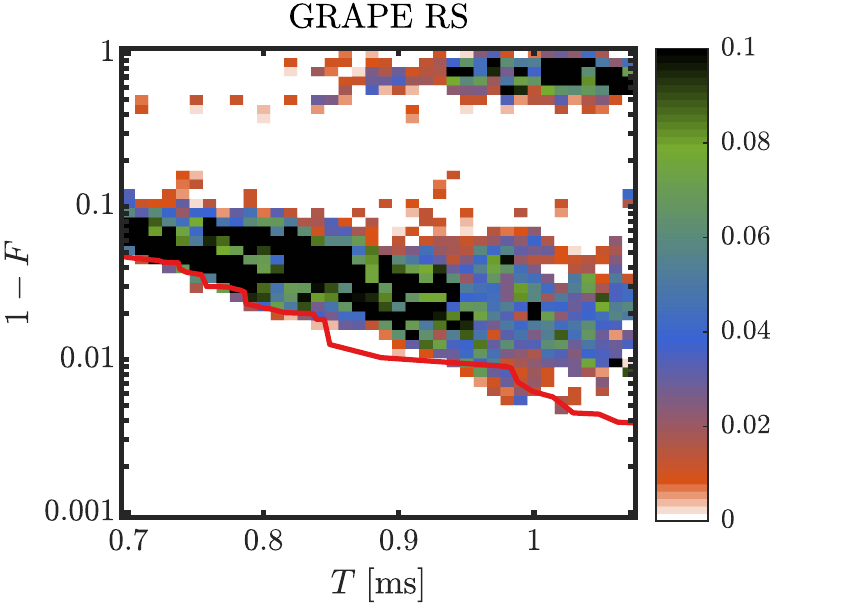} \\
			\includegraphics[]{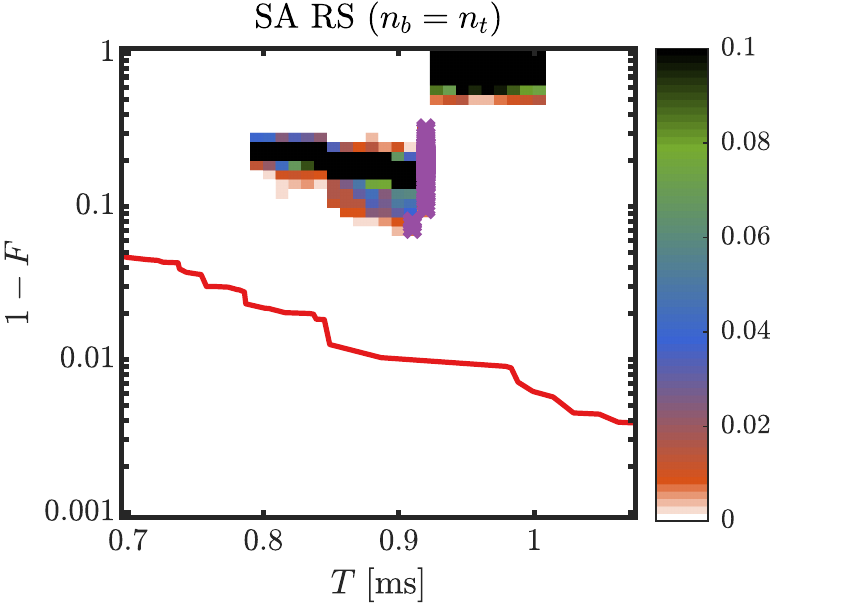}
			\includegraphics[]{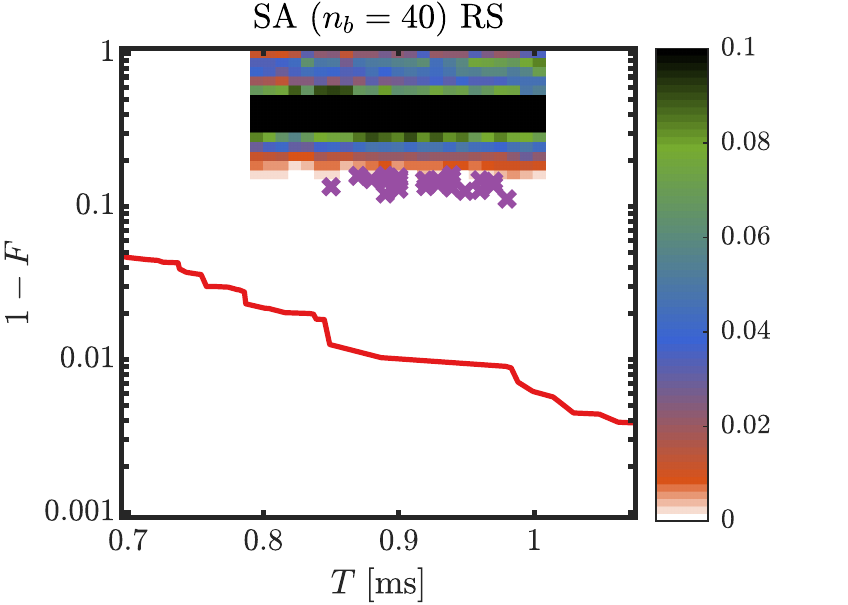} 
		
		\caption{\textit{Shake Up} solution densities for different methods. Each density is normalized for every individual $T$ and thus represents an estimate of the probability distribution for obtaining a particular $F$ for a given $T$ ($\mathcal{P}(F|T)$). The red reference curve shows the best obtained results for \grape \ps. Purple crosses indicate individual solutions at densities lower than 0.002.
		}
		\label{fig:solutiondensities_shakeup}
		\hspace{-1.75cm}
				\includegraphics[]{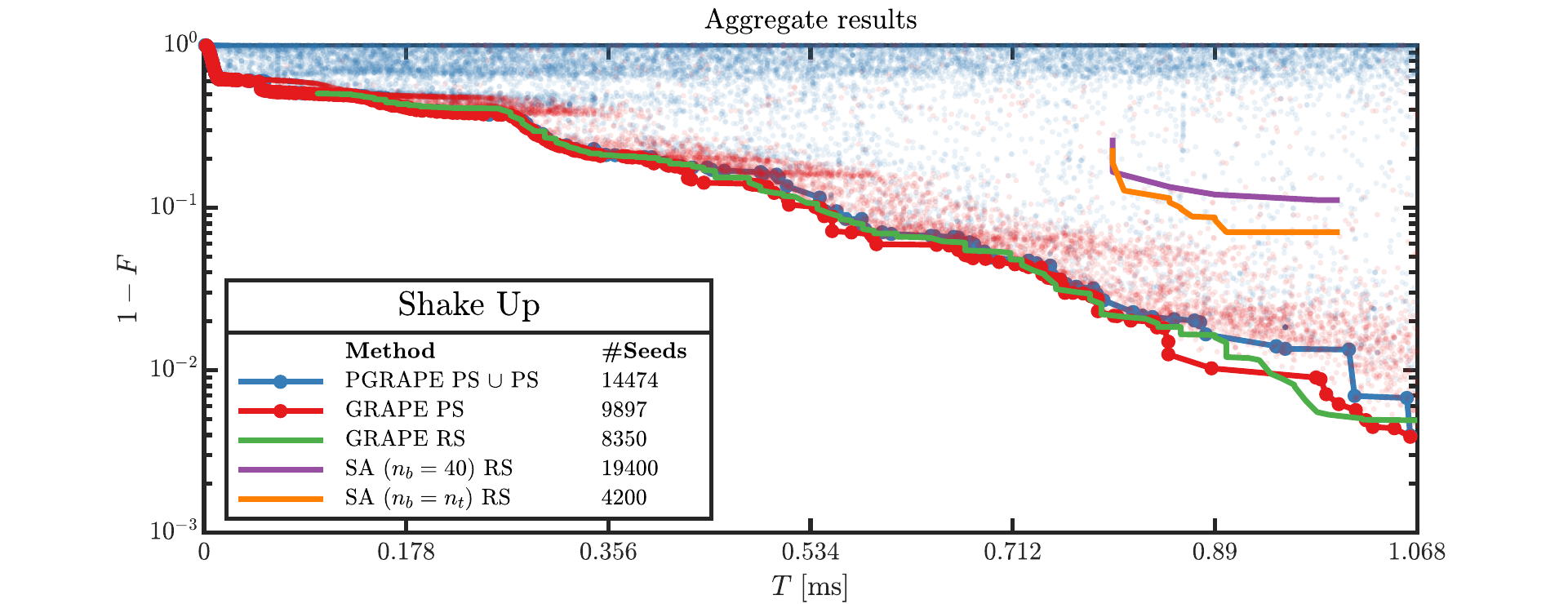}

		\caption{Aggregate, monotonically best optimization results (lower is better) for several methods in \textit{Shake Up}. 
			Solid lines with (without) dots show the best results obtained with (without) player influence.
			The scattered blue dots show all results produced by 266 players seeding on average $\sim 37$ solutions and optimizing approximately $1/3$ of these for $\sim 116$ iterations on average.  
			The scattered red dots show the same, except the optimization is carried out on all player seeds with the computational resources described in the text (i.e. no player influence after seeding). The dot translucency indicates the density distribution. 
		}
		\label{fig:FT_shakeup}
	\end{minipage}
\end{figure*}

The solution densities are shown in Fig.~\ref{fig:solutiondensities_shakeup} and the aggregate results are shown in Fig.~\ref{fig:FT_shakeup}. 
The kinks in the reference curve in Fig.~\ref{fig:solutiondensities_shakeup} can be more apparently understood by looking at the red dots in Fig.~\ref{fig:FT_shakeup} (corresponding to \grape \ps solutions). 
The several pronounced, staircase-like plateaus suggest the existence of multiple strategies that are relevant at different duration intervals. 
These are examined more closely in Sec.~\ref{sec:clustering_shakeup} and are found to be associated with solution strategies defined by elements of periodic modulation.
Each plateau extends over the next, meaning that the now inferior, locally optimal strategy remains a prominent attractor for quite some interval of duration. 
From looking at the \grape solution densities, this evidently makes the new globally optimal strategy much harder to find. 
For \grape \ps the density splits when crossing the kinks and one plateau in particular remains the main attractor, obfuscating the globally optimal strategy that coincidentally exists in the wing of the distribution. 
On the other hand, the density

\clearpage
\makeatletter\onecolumngrid@push\makeatother

\clearpage
\clearpage
\makeatletter\onecolumngrid@pop\makeatother

\noindent of very low fidelity solutions is sparse and independent of $T$. 
This occurs because partial transfers are not difficult to achieve as the initial state density can very quickly and easily be overlapped with one of the lobes of the double-peaked target state density (see Fig.~\ref{fig:levels}) by a small constant displacement.
Looking at the aggregate results, we indeed observe that the red dots are separated from the upper, thick sea of blue dots. The abundance of the latter at low-fidelity is due to players terminating their optimization prematurely.

No single method is the best for all $T$. 
This is contrary to both \textit{Bring Home Water} and \textit{Splitting}, where \grape \ps,  \pgrape \ps $\cup$ \ps and either \sa \rs or \grape \rs \,\footnote{
	For the \textit{Shake Up} analysis we had initially produced a factor of 6 more \rs seeds than \ps seeds. To make the comparison fair, we form subsets by discretizing the $T$ axis into 24 bins and randomly drawing \rs seeds equal to the number of \ps seeds in each bin (the ensuing difference in total number of seeds in Fig.~\ref{fig:FT_shakeup} is due to the lack of \rs seeds at very short durations). The presented solution density and aggregate results are representative of the statistical outcome of this procedure.}, respectively, found the globally optimal solutions independently. In \textit{Shake Up}, however,  \ps seeds seem to have the upper hand around $T = 0.85\,\si{ms}$ and after $T = 0.98 \,\si{ms}$, while the \rs seeds seem to be better between those durations. 
The \grape \rs density provides some nuance to this observation. It shows a broader, less dense distribution of high-fidelity solutions and the addition of many very low fidelity solutions as $T$ increases. In fact, near $T=1.05\,\si{ms}$ the monotonically best results are due to only a few points in an otherwise empty density region. On the contrary, for \grape \ps, the same region has a fairly high density and is thus the statistically superior method in the $T=1.05\,\si{ms}$ regime. We provide another perspective on these statistical performances in Sec.~\ref{sec:sampling}.
This gives rise to slightly different quantum speed limit estimates:
$T\st{QSL}^{F=0.99} \approx 0.939\,\si{ms}$ for \rs and $T\st{QSL}^{F=0.99} \approx 0.969\,\si{ms}$ for \ps, although the point at $T=0.887\,\si{ms}$ with $F=0.9897$ comes very close to tipping this (somewhat arbitrary) balance. Neither method obtained an estimate for $T\st{QSL}^{F=0.999}$ within the span of durations present in the game. 
The overall behavior of \pgrape \ps $\cup$ \ps shows that players terminated the optimization prematurely in this problem and therefore yields an approximately $7\%$ worse estimate.



For the \sa \rs methods, the $(n_b = n_t)$ and $(n_b = 40)$ variants both fail to find any meaningful results. Whereas the \textit{Splitting} control landscape was benign enough to compensate for the computational difficulties associated with $g\neq 0$, this is clearly not the case in \textit{Shake Up}.

\subsection{Optimal Strategies -- Control Clustering}
\label{sec:clustering_shakeup}
For \textit{Shake Up}, the optimized controls did not possess any readily apparent structure. When individually plotted alongside the corresponding position expectation value of the wave function, however, 
an oscillatory structure begins to emerge. Subtracting the expectation values, $u(t) - \braket{x(t)}$, we observe that these relative controls are
dominated by low frequency cosine components. 
Decomposing the relative controls as 
\begin{align}
c_k = \frac{1}{T} \int_{0}^{T} (u(t) - \braket{x(t)}) \cos(\pi k  t/T) \d t 
\end{align}
thus yields low-dimensional vectors $\vec c = (c_0, \dots, c_5)$ in frequency space (with corresponding number of oscillations $N_k = k/2$) on which we apply clustering. 
We join the \grape \ps and \grape \rs result sets, selected by the criteria $0.267 \,\si{ms} < T < 1.068\,\si{ms}$ and $F > 0.6$,
since the player and random seeds were dominant in different regions. We use $\epsilon = 0.1$ and $\mathrm{min}\st{samples} = 250$. These parameters were chosen such that a single cluster is identified per cosine component. 

\begin{figure}

				\includegraphics[width=\linewidth]{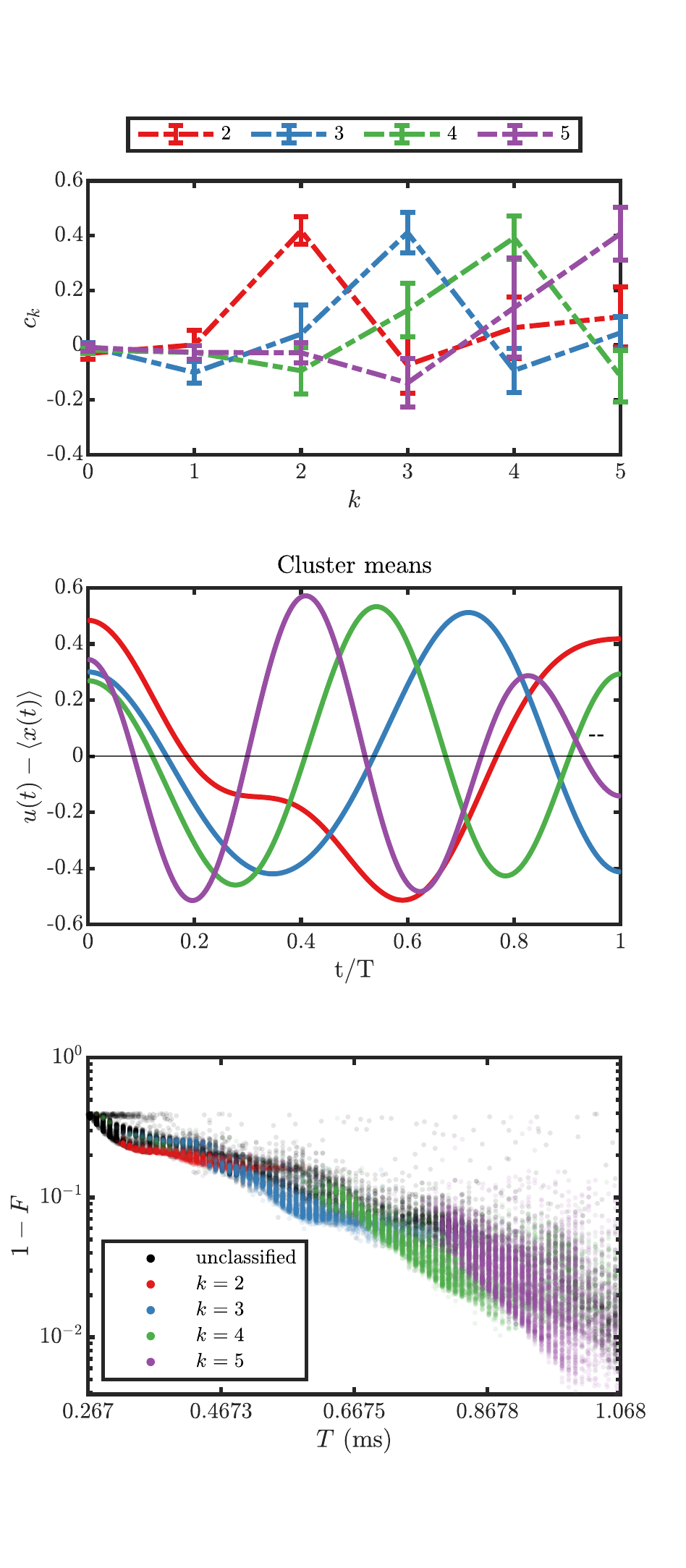}
	\caption{\textit{Shake Up} clustering results. Top: Component weights $c_k$, with colors corresponding to the different clusters. The dots with bars denote the mean and standard deviation, whereas the dashed lines only help to guide the eye.
	Middle: Cluster means in real space color-coded as above. Bottom: Fidelity distribution color-coded as above.}
	\label{fig:shakeup_clustering}
\end{figure}

Figure~\ref{fig:shakeup_clustering} shows the clustering results. Each cluster is labeled by its dominant coefficient corresponding to a half-integer or integer number of oscillations, which is evident from the cluster means when transforming back into real space. 
When solutions are colored according to their cluster membership in the aggregate plot, we uncover a clear hierarchy of the solution strategies corresponding to the clusters: 
within the full set of time intervals considered, each strategy is sequentially globally optimal in ascending order of oscillation number with approximately equal interval lengths.
As the globally optimal strategy transitions, the now inferior locally optimal strategy remains a relatively broad attractor for an appreciable interval of duration $T$, leading to the plateaus observed in Fig.~\ref{fig:FT_shakeup}.

\medskip
Based on the presented analysis, we do not believe we have found the true best results and associated quantum speed limits since (i) neither \pgrape \ps $\cup$ \ps, \grape \ps, or \grape \rs alone produce the best results, (ii) the solution densities are shifted away from the optimal strategies with high variances, and (iii) \sa fails completely for this task.
In particular, the \grape \ps solution at $0.887\,\si{ms}$ in Fig.~\ref{fig:FT_shakeup} strongly suggests the existence of optimal solutions in this vicinity from the $k=4$ solution strategy shown in Fig.~\ref{fig:FT_shakeup}. Their discovery, however, is evidently obfuscated due to the other active, locally optimal solution strategies ($k=3,5$) in this region. Based on this analysis one could imagine alleviating this issue by developing a seeding mechanism parametrized to specifically target the $k=4$ strategy. 

This demonstrates that this is the most challenging of the three examined problems in terms of landscape complexity.
However, we have successfully identified the hierarchy of solution strategies that could guide further attempts at locating the ultimate quantum speed limits.

\section{Statistical Performance}
\label{sec:sampling}

Here, we provide an alternative way of characterizing the statistical method performance, which is not based on solution densities, and in addition assess the relative usefulness of the preselection heuristic introduced in Sec.~\ref{sec:algseed}.
Instead, we compare statistics when only a restricted number of seeds are allowed to be optimized. 
This emulates either the restriction of computational resources (e.g. no cluster for parallel computation is available) or an increased numerical difficulty of the problem (i.e. operations are more expensive in wall time, allowing fewer seeds to be optimized within the same time frame).

Regardless of either interpretation, we estimate a mean quantum speed limit $\braket{T\ut{fit}\st{QSL}}$ for $F=0.99$ as a function of the sample size, $N\st{samples}$, on the aggregate results in Figs.~\ref{fig:FT_bhw}, \ref{fig:FT_split}, and \ref{fig:FT_shakeup}. 
The procedure is based on fitting randomly sampled solutions in a specified interval $\mathcal T_\sqcup$. Concretely, the procedure runs as follows:


\begin{enumerate}[itemindent=\theitemindent em,align=left,leftmargin=*]
	\item Divide the interval $\mathcal{T}_\sqcup = [0.8,1.2] \cdot T\st{QSL}^{F=0.99} $ into 15 sub-intervals of equal width. 
	\item Sample $N\st{samples}$ solutions within $\mathcal T_\sqcup$. 
	\item Within each sub-interval, select the solution with the highest fidelity and discard the rest.
	\item Linearly fit the subset from step 3 with $\log_{10}(1-F(T))$, i.e. assume exponential behavior in $T$, and denote by $T\ut{fit}\st{QSL}$ the duration where the fit value corresponds to $F=0.99$. 
	\item If $T\ut{fit}\st{QSL} \in  \mathcal T_\sqcup$ (interpolation) it counts as a success. If it lies outside $\mathcal T_\sqcup$ (extrapolation) it counts as a failure. 
	\item Repeat 2-5 $N\st{trials} = 1000$ times.
	\item Compute the mean value $\braket{T\ut{fit}\st{QSL}}$ over successful trials.
	\item Compute the empirical success rate (estimated probability of success) $\mathcal{P} (T_\mathrm{QSL}^{^\mathrm{fit}} \in \mathcal T_\sqcup)  =  N\st{successes} / N\st{trials}$.
\end{enumerate}
Including only the successful trials in $\braket{T\ut{fit}\st{QSL}}$ avoids skewing the mean due to extreme outliers (i.e. due to too small negative slope), and the information about these is instead captured in $\mathcal{P} (T_\mathrm{QSL}^{^\mathrm{fit}} \in \mathcal T_\sqcup)$. 
Thus, for a randomly sampled set of solutions, their $T\ut{fit}\st{QSL}$ will in the mean be $\braket{T\ut{fit}\st{QSL}}$ with probability $\mathcal{P}$, or fail with probability $1-\mathcal{P}$. 
In the following we denote triples of these quantities as $\{N\st{samples}, \braket{T\ut{fit}\st{QSL}},\mathcal{P}\}$.

Fig.~\ref{fig:sampling} shows the results for different methods in all three levels. 
Taking the generic \grape \rs method as a baseline comparison, one finds that the average behavior of
\pgrape \ps $\cup$ \ps (in-game player seeding and player optimization) tends to be worse in all levels: it has a comparatively high $\braket{T\ut{fit}\st{QSL}}$ and a low 
$\mathcal{P}$ for small sample sizes, for instance  $\{30, 1.15, 0.2\}$ in \textit{Bring Home Water}.
An exception occurs beyond $N\st{samples} = 200$ where the players in-game perform better on average, but only in this level.

Thus player methods without \textit{additional} optimization did not yield a superior approach on average. 
Upon optimization, however, \grape \ps displays significant statistical performance increases over \grape \rs across all three levels (implying also that players terminated their optimization prematurely and thus did not themselves realize the full potential of their seeds).
This is particularly illuminating for \textit{Shake Up} as neither of the two methods was clearly shown to be better in Sec.~\ref{sec:shakeup}. There, $\mathcal{O}(10^3)$ seeds were optimized, and at those sample sizes in Fig.~\ref{fig:sampling} we indeed find coincidence of the two methods' statistical performance capacities. However, as $N\st{samples}$ is decreased, \grape \ps becomes statistically superior. \medskip

We now turn to the effectiveness of the preselection heuristic. 
Preselecting (\pr) the best individual 600 \ps and \po (i.e. the output from \pgrape \ps) seeds and optimizing these we see a significant shift in statistical performance.
The \grape  \pr-\po and \grape \pr-\ps methods are observed to require up to several orders of magnitude fewer samples to produce the same $\braket{T\ut{fit}\st{QSL}}$ across all three levels compared to the baseline \grape \rs (without preselection). Moreover, their success rates are significantly higher than any other method for small sample sizes (rivaled only by \sa in \textit{Bring Home Water}) and only dips below unit $\mathcal{P}$ in \textit{Shake Up}. 
For the most extreme case, compare $\{30, 1.135, 0.95\}$ for \grape \pr-\po and $\{30,  1.146, 0.07\}$ for \grape \rs.
Obtaining the same $\braket{T\ut{fit}\st{QSL}}$ and $\mathcal P$ in \grape \rs occurs at much higher sample sizes ($\{120, 1.135, 0.70\}$, and $\{260, 1.109, 0.95\}$, respectively). 
Similar trends are seen for the other levels, e.g. in \textit{Bring Home Water} at $N\st{samples}=30$, where \grape \pr-\po obtains $\{30, 0.999, 1\}$ and \grape \rs obtains $\{30, 1.104, 0.688\}$.

Based on these findings, it might be reasonable to expect that preselection of \rs seeds would lead to similar improvements.
In this case, however, much smaller relative improvements are observed.
This shows that, across all levels, the structure of the best player seeds places them much more prominently in the (abstract) green region of optimality of Fig.~\ref{fig:landscape} than the best random seeds. 

The $\braket{T\ut{fit}\st{QSL}}$ reduction gained by increasing the sample size for \grape \pr-\po is minimal and quickly saturated in \textit{Bring Home Water} and \textit{Splitting}, while this is not true for \textit{Shake Up}. This reinforces the conclusion that \textit{Shake Up} is the most difficult problem overall considered in this work.
The only instance where \grape \pr-\po is matched in performance at the same $N\st{samples}$ occurs in \textit{Splitting}.
Here \grape \pr-\rs intersects near $N\st{samples}=200$, the maximum sample size for the preselection-based methods. 
However, practically the same $\braket{T\ut{fit}\st{QSL}}$ can be achieved using just $N\st{samples}=30$ with \grape \pr-\po.

Additionally, the relative difference between \pr-\po and \pr-\ps signifies roughly how valuable the in-game player optimizations were in terms of absolute results for a given level (cf. statistics in captions of Figs.~\ref{fig:FT_bhw}, \ref{fig:FT_split}, and \ref{fig:FT_shakeup}). 
\textit{Bring Home Water} and \textit{Shake Up} show a clear gain while \textit{Splitting} is nearly unaffected, lending itself again to the interpretation that it has the least difficult landscape topography. 

The \sa ($n_b = n_t$) method performs very well in \textit{Bring Home Water} (with the caveats described in Sec.~\ref{sec:bhw} also discussed shortly). The \sa ($n_b = 40$) method is unsurprisingly seen to have the unequivocally worst overall statistical performance, even in the linear case, $\braket{T\ut{fit}\st{QSL}}$ scales comparatively poorly with $N\st{samples}$. For \textit{Shake Up} both these methods fail ($\mathcal{P} = 0$) for all $N\st{samples}$. The failure is, again, less severe for \textit{Splitting}.
\medskip

\subsection{Algorithmic Run Time}
Observations on end-of-optimization results in Figs.~\ref{fig:FT_bhw}, \ref{fig:FT_split}, and \ref{fig:FT_shakeup} do not account for the associated computation times, except in the time-out termination condition.

\begin{figure*}[b]
	\begin{minipage}{\textwidth}
		\centering
		\pgfmathsetmacro{\theleftcrop}{1.2}
		\pgfmathsetmacro{\theleftspace}{0.05} 
		\pgfmathsetmacro{\therightcrop}{0.0}
		\pgfmathsetmacro{\thebottomcrop}{3.8}
		\pgfmathsetmacro{\thetopcrop}{1.8}
		
			\includegraphics[,trim={0cm \thebottomcrop cm \therightcrop cm \thetopcrop cm},clip]{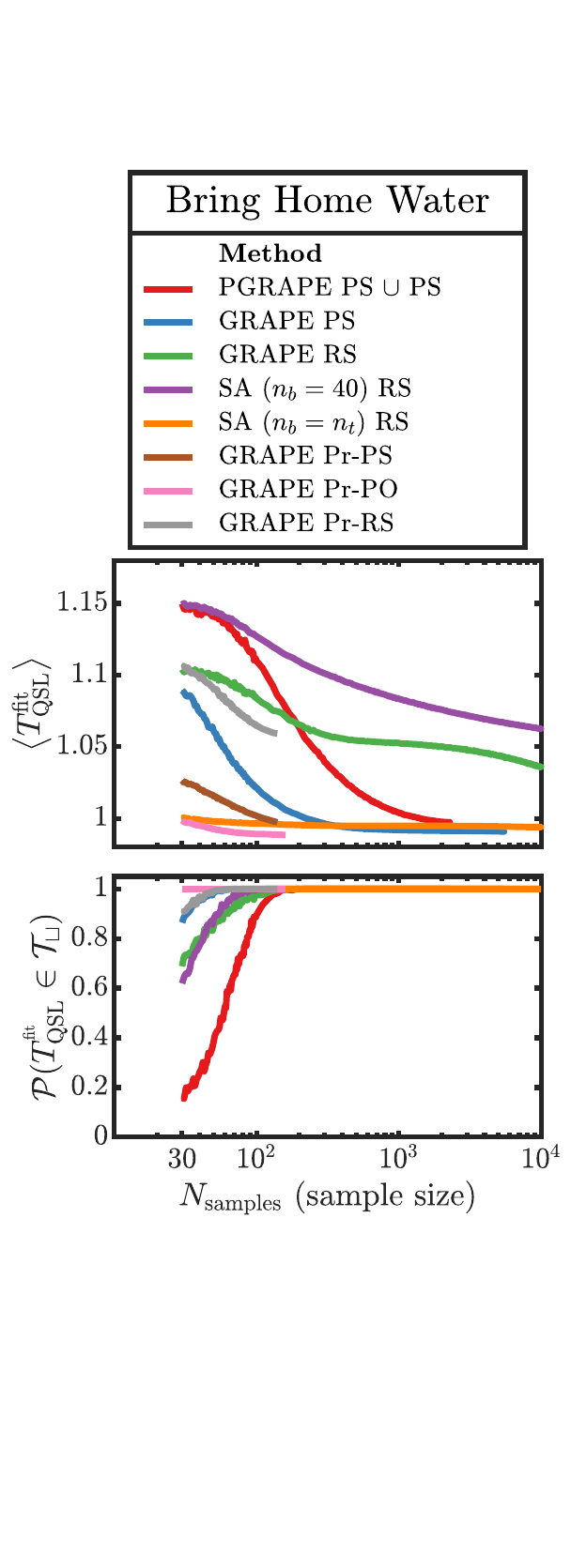}
			\hspace{\theleftspace cm}
			\includegraphics[,trim={\theleftcrop cm \thebottomcrop cm \therightcrop cm \thetopcrop cm},clip]{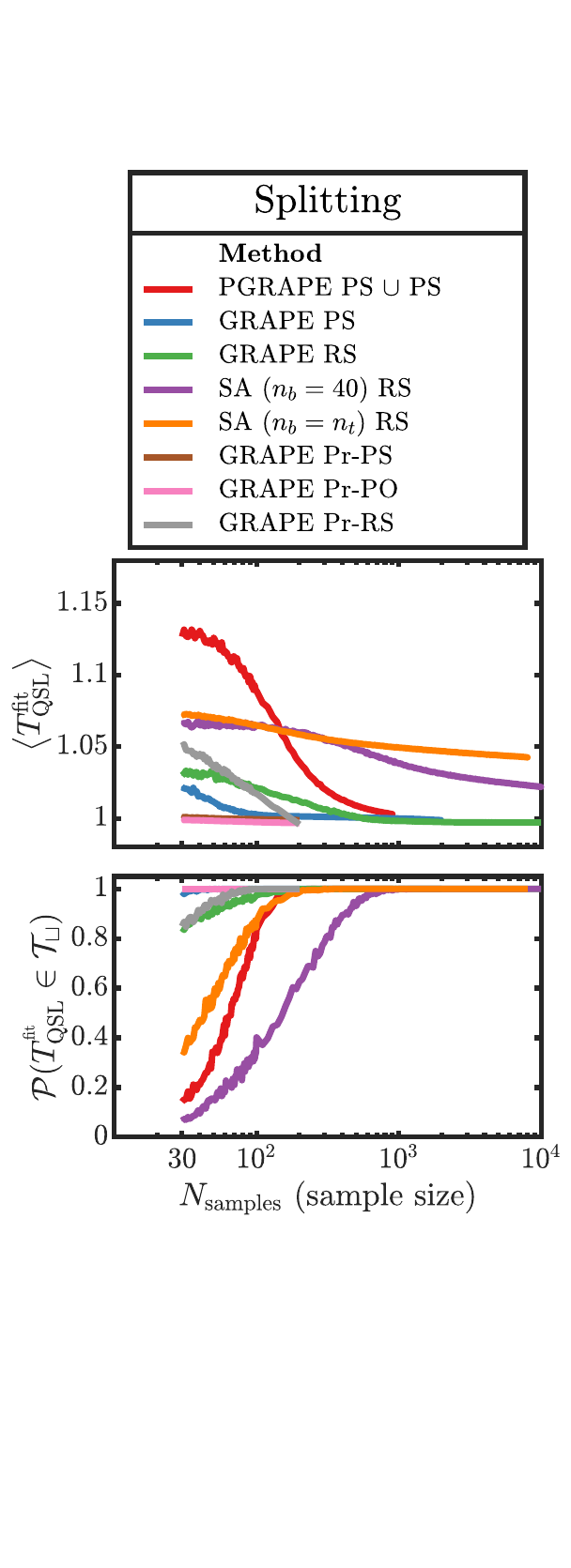}
			\hspace{\theleftspace cm}
			\includegraphics[,trim={\theleftcrop cm \thebottomcrop cm \therightcrop cm \thetopcrop cm},clip]{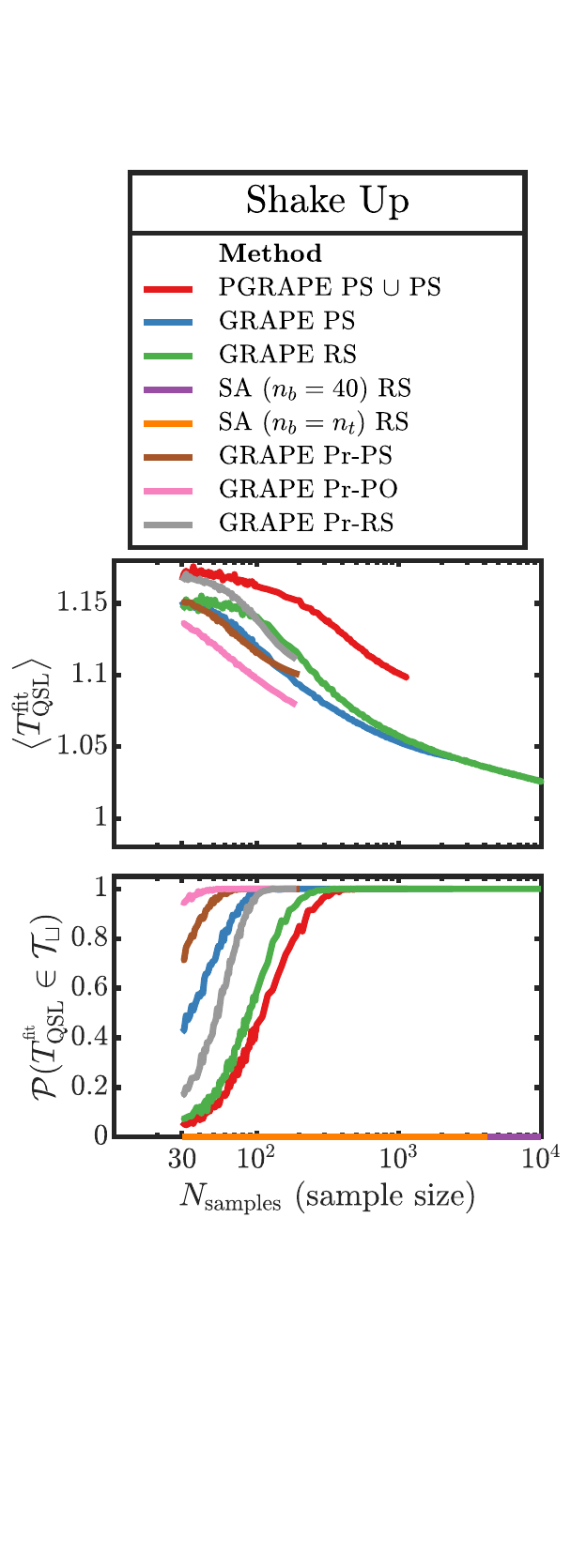}
		\caption{Top row: Estimated $\braket{T\ut{fit}\st{QSL}}$ in units of ${T^{F=0.99}\st{QSL}}$.
			Bottom row: Empirical success rate (estimated probability of success) $\mathcal{P} (T_\mathrm{QSL}^{^\mathrm{fit}} \in \mathcal T_\sqcup)$ as a function of sample size.
			Strong statistical performance is signified by low $N\st{samples}$ having simultaneously low $\braket{T\ut{fit}\st{QSL}}$ and high $\mathcal{P}$. 	
	}
		\label{fig:sampling}		
		\pgfmathsetmacro{\theleftcrop}{0.8}
		\pgfmathsetmacro{\theleftspace}{-0.2} 
		\pgfmathsetmacro{\therightcrop}{0.0}
		\pgfmathsetmacro{\thebottomcrop}{0}
		\pgfmathsetmacro{\thetopcrop}{0}
		
		\hspace{0.22 cm}
		\includegraphics[,trim={0cm \thebottomcrop cm \therightcrop cm \thetopcrop cm},clip]{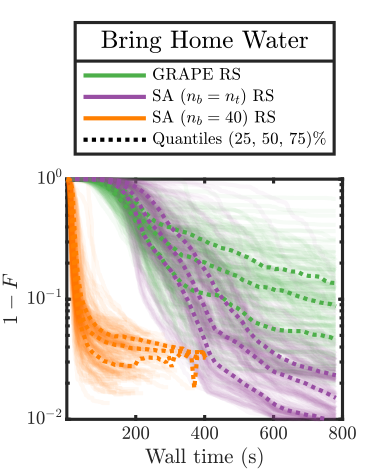}
		\hspace{\theleftspace cm}
		\includegraphics[,trim={\theleftcrop cm \thebottomcrop cm \therightcrop cm \thetopcrop cm},clip]{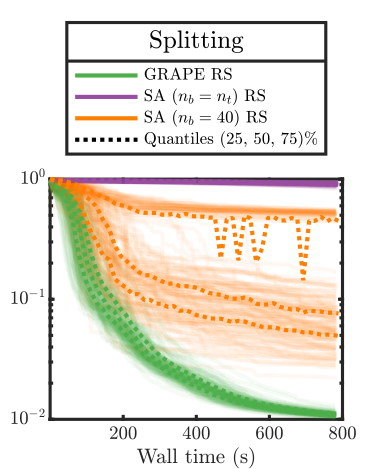}
		\hspace{\theleftspace cm}
		\includegraphics[,trim={\theleftcrop cm \thebottomcrop cm \therightcrop cm \thetopcrop cm},clip]{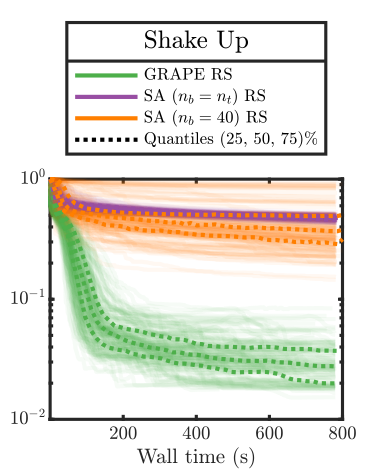}
			\caption{Results (lower is better) for \grape (green), \sa $(n_b = n_t)$ (purple), and \sa $(n_b = 40)$ (orange) using the same 100 \rs seeds at $T\st{QSL}^{F=0.99} \approx 0.0973,  0.92, 0.89 \,\si{ms}$ for the three levels, respectively. 
			Each line denotes an optimization trajectory as a function of wall time. The translucency is related to the density distribution and the three dotted lines indicate the $25\%, 50\%,$ and $75\%$ quantiles for all active optimizations. 
		} 
		\label{fig:SAvGrape_bhw} 
	\end{minipage}
\end{figure*}

Figure~\ref{fig:SAvGrape_bhw} shows the optimization trajectories for \rs methods as a function of wall time 
for the same 100 seeds 
at or near $T\st{QSL}^{F=0.99}$ in each level
\footnote{Comparing iterations is not very instructive since the algorithms are so fundamentally different.}.  
These are separate from the previous results, and each seed had in this case a maximum optimization time of roughly 13 minutes.
The quantile statistics consider only ongoing optimizations, i.e. they do not include solutions converged at earlier times.

In \textit{Bring Home Water}, we clearly see that \sa ($n_b = 40$) rapidly improves the fidelity initially, lending merit to its usefulness (as discussed in Appendix~\ref{app:algorithms}). This is owed to the speed at which a full iteration can be carried out, i.e. where all parts of the control domain are adjusted once.
The mean time per complete iteration is $14.2 \pm 1.4\,\si{s}$. Further speed up can be achieved as pointed out in Appendix~\ref{app:algorithms}.
However, progress stagnates and terminates before $F=0.99$ due to the reduced resolution.
A better choice for the heuristic resolution parameter $40 < n_b < n_t$ would address this issue. 
The full resolution \sa ($n_b = n_t$) does not suffer from stagnation, but progress is initially orders of magnitudes slower since control values at successive times are completely uncorrelated and a full iteration takes much longer. The mean time per complete iteration is $219 \pm 31\;\si{s}$. 
\grape completes an iteration in $0.27 \pm 0.03\,\si{s}$ and performs about the same as \sa ($n_b = n_t$) until around 400s. Beyond this point \sa ($n_b = n_t$) dominates in the mean, reflecting that \grape \rs finds only the inferior, locally optimal strategy with less efficiency.
Recall that, as opposed to \grape, \sa optimizes only $u_1(t)$ and is not subject to derivative regularization (smoothness criterion) \cite{sorensen2019qengine}  of the control, and the algorithm therefore effectively solves an easier problem in the current implementation. Inclusion of both points would likely lead to a doubling in computation time (i.e. stretching the purple and orange lines by a factor 2 on the $x$-axis in Fig.~\ref{fig:SAvGrape_bhw}) in order to obtain similar results to the ones presented, making them more or less coincidental with \grape \rs near $800\,\si{s}$. However, we do not believe this would drastically change the conclusions in terms of the overall performance in Sec.~\ref{sec:bhw}.

In \textit{Shake Up}, \sa ($n_b = 40$) and  \sa ($n_b = n_t$) complete an iteration in $159 \pm 13\,\si{s}$ and $>800\,\si{s}$, respectively, but never reach high fidelities. \grape completes an iteration in $0.76 \pm 0.08\,\si{s}$ and fares comparatively much better. Similar numbers 
are found in \textit{Splitting}.
%
%

\clearpage

\makeatletter\onecolumngrid@push\makeatother

\clearpage
\makeatletter\onecolumngrid@pop\makeatother

\section{Discussion and Outlook}
\label{sec:conclusion}
We have presented \textit{Quantum Moves 2}, a citizen-science game in which players act as seeding mechanisms and initial optimizers for quantum optimal control problems.
Selecting three distinct problems in the game for analyses, we applied different optimization methods (combinations of algorithms and seeding strategies) to these. For each problem, we examined the (a) respective methods' results and efficiencies, (b) optimal solution strategies, (c) overall problem structures, and (d) statistical performance capacities through a separate random sampling procedure. Here we summarize the findings. \medskip

In \textit{Bring Home Water}, a single-particle problem, we identified two solution strategies (front- and back-swing) characterized by an exponentially-widening gap between them. Using the same resources and a gradient-based algorithm (\grape), the player-infused seeds uncover both strategies efficiently whereas the random seeds only find the inferior, locally optimal strategy at durations relevant for high-fidelity transfers. Imposing increasingly more structure \textit{a posteriori} from physical insight on the random seeds allows discovery of the globally optimal strategy, albeit with significantly reduced overall efficiency. 
Employing a gradient-free stochastic ascent algorithm (\sa) with random seeds and a strong enforced behavior on one of the controls (tweezer maximally deep at all times), we find both strategies with high efficiency. 
The success of \sa is explained by a combination of the optimization landscape's structure (many, small local traps and a few, broad optimal attractors),  
the algorithm's search methodology (allowing it to escape the abundant local traps), and the linearity of the equations of motion (reducing algorithmic complexity). On an optimization algorithmic level, the comparatively reduced efficiency of the gradient-based algorithm is due to its inability to escape these traps. 
Thus, the efficiency is due to the algorithm and not the seeding strategy; using \sa with player seeds is expected to yield results at least as good as \sa with random seeds.

In \textit{Splitting}, a \bec problem, we identified a single solution strategy. This strategy was found efficiently by both the player-infused  and randomly seeded gradient-based methods. The gradient-free algorithms with random seeds fail due to the non-linearity of the equations of motion (resulting in increased algorithmic complexity), with the exception that one variant is somewhat successful in the limit of large durations. Here, the optimization landscape is thus so simple that even a numerically inefficient algorithm can discover the globally optimal strategy with the allotted resources. 

In \textit{Shake Up}, another \bec problem, we identify four solution strategies involving low frequency oscillations around the \bec center-of-mass. Each strategy, individually characterized by a dominant half-integer number of oscillation periods, is globally optimal at different durations in order of lowest to highest number of oscillations. Outside of their respective regions, each strategy remains a broad locally optimal attractor, leading to plateaus in the optimization results. Based on these results, neither of the gradient-based methods are found to be the best on their own, but each exhibits exclusive regions of dominance. In this instance, the gradient-free algorithms completely fail. From these observations, we conclude that the optimization landscape is very complex and we do not believe we have found the ultimately best results. 

Through random sampling, we then found that the statistical performance of player seeds was always better than random seeds in all the studied levels when restricting the sample size (equivalent to increased numerical difficulty or restricted available computational resources). This could be considered a tiebreaker for the methods' similar absolute performances in the \textit{Shake Up} problem. 
Additionally, preselecting the player solutions based on best initial fidelity significantly increases their relative performance when optimized, whereas the same preselection procedure improves the random seeds to a lesser comparative extent. 
This can be interpreted as the best player solutions being more likely to be located in the green region of optimality in Fig.~\ref{fig:landscape}.
The benefit diminishes gradually as the sample size is increased and the overall optimization landscape more densely explored.
This echoes previously drawn conclusions -- given enough resources (correspondingly a sufficiently large sample size) most combinations of algorithms and seeding strategies achieve similar absolute results, given they at least partially cover the green region of optimality.

\medskip

\medskip
\medskip

We now return to the two citizen science related questions discussed in the introduction. A main feature of \textit{Quantum Moves 2} was an in-game optimization button enabling players to store and optimize candidate solutions on their local device. This clearly underscored the player’s role in the search for overall, global features in solutions (that is, locating the green region of optimality in Fig.~\ref{fig:landscape}), whereas fine-grained, local optimization could be left to the optimization algorithm. This certainly supports a sequential, ``one-off'', player-computer interaction. However, with supporting tools (such as replay and the \textit{ghost} feature, see Appendix~\ref{app:game}) the game also enables more intertwined,
hybrid human-computer interactions in which players gain insight by examining the output of the computer optimization and can thereby improve their search for promising features and heuristics. Further study of this will be left for future work.

Figures \ref{fig:FT_bhw}, \ref{fig:FT_split}, and \ref{fig:FT_shakeup} demonstrate that
the method of player-seeding with player-invoked local-device optimization (\pgrape \ps $\cup$ \ps)
performs roughly on par (in terms of best achieved results) with the best performing fully algorithmic approaches under consideration in each problem. For the two hardest challenges, \textit{Bring Home Water} and \textit{Shake Up}, this method outperforms the randomly seeded \grape and \sa, respectively. Thus, we conclude, as proposed in Q1, that it does indeed make sense to develop a framework, like Foldit has done for protein folding, in which the solution of quantum problems are outsourced to the general population.
In Q2, we ask if the game-based approach could actually yield a computational advantage. 
A full answer to this question should entail comprehensive comparison to the best possible expert-driven optimization. Here, we take first steps in that direction with a baseline benchmark comparison to off-the-shelf optimization with initialization that is as heuristic-free as possible.
We find that \grape with player seeds (\grape \ps) is the only method that is roughly optimal across all three levels, the exception being a small window in \textit{Shake Up}. These results should only be understood as a necessary baseline study and a first demonstration for further exploration, and they should not be taken as a guarantee that player-based seeding is advantageous when comparing to increasingly complex algorithmic strategies. However, outside of immediate performance capabilities, player-generated data may show additional potential because, similar to machine learning-generated data 
\cite{mogens2020global,iten2020discovering,krenn2016automated}, they constitute a means for researchers to address problems that is not influenced by any expert biases. This can inform the extraction of heuristics and insights that can subsequently be understood, utilized, and expanded upon by the domain expert. 
An indication of this was seen in the most complex challenge, \textit{Shake Up}, in which player and randomly seeded methods were globally optimal at different durations, as they probed different parts of the interleaved optimal strategies. 

Based on the answers to Q1 and Q2, we suggest posing additional questions for this line of research:
\begin{enumerate}[label=Q\arabic*:]
	\setcounter{enumi}{2}
	\item  Could citizen science games be first steps towards playful, explorative tools for domain experts? This approach is currently being pursued in microbiological research settings \cite{cooper2018repurposing} but not yet in quantum physics. As a related question, can the citizen science experience be expanded to include citizen contribution in more aspects of the scientific method such as data analysis and, ultimately, hypothesis and problem formulation? If such so-called extreme citizen science \cite{haklay2013citizen} could be understood and systematically implemented then this would constitute a major advance 
	 in complex problem solving, one of the most demanded skills as per the World Economic Forum \cite{wef2018}. 
	\item If the games are sufficiently challenging for humans, can they be used for systematic studies of human problem solving? In our group we have started to investigate this within the setting of quantum experimental optimization \cite{heck2018remoteopt} and by developing cognitive science variants of quantum challenges \cite{quantumminds}.
	\item If larger portions of the player base can make non-trivial contributions to several classes of research problems using the gamified interface, could this  contribute to the solution of one of the major roadblocks in the path to domain-general AI according to e.g. the author of Ref.~\cite{marcus2020next} - that is - the crowdsourcing and algorithmization of human common sense?
\end{enumerate}
The latter two questions move well beyond the realm of quantum physics and the scope of this paper. We currently pursue Q3 in other work \cite{ahmed2020quantum} by developing an intuitive and visual quantum programming environment. 
The extent to which our work may be extended to gain broader implications for the fields of quantum research (Q3), social science (Q4) and computer science (Q5) constitutes interesting topics for future studies.

\section{Acknowledgements}
We acknowledge support from European Union's Horizon 2020 research and innovation programme under the Marie Sklodowska-Curie QuSCo grant agreement N$^\mathrm{o}$ 765267, the ERC, H2020 grant 639560 (MECTRL), and the John Templeton and Carlsberg Foundations.
The numerical computer cluster optimization results presented in this work were obtained at the Centre for Scientific Computing, Aarhus \url{phys.au.dk/forskning/cscaa}.


\appendix

\clearpage
\section{Interface}
\label{app:game}
\begin{figure}[h]
	\begin{center}

			\includegraphics[width=1\linewidth]{fig_splitting_level}
			\\
			\vspace{0.1cm}
			\includegraphics[width=1\linewidth]{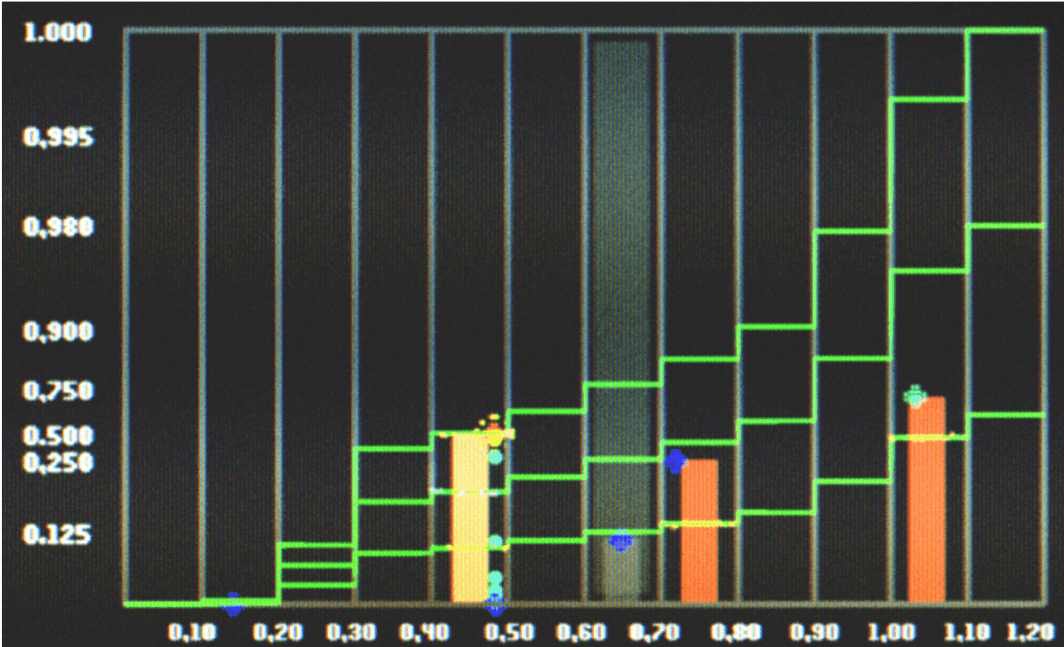} 
			\\
			\vspace{0.1cm}
			\includegraphics[width=1\linewidth]{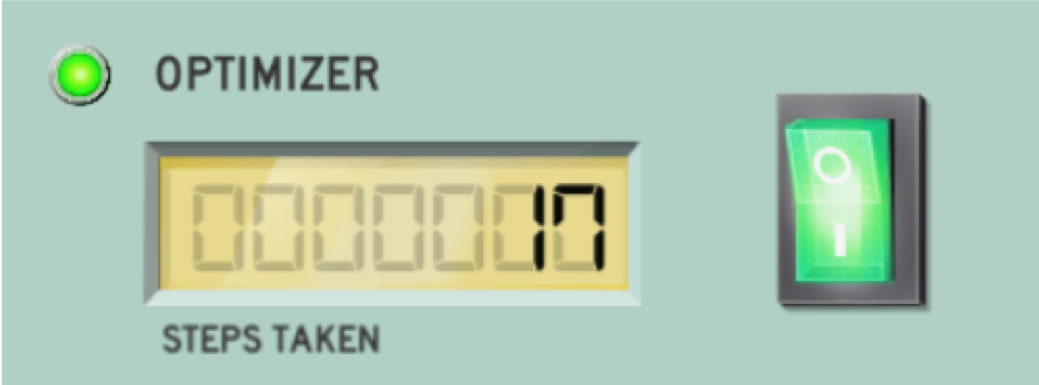}
	
		\caption{Partial screen shots of the two main views in \textit{Quantum Moves 2}. Much of the UI has been hidden for simplicity.	
			Top: Dynamic play view with dynamic potential (green), instantaneous wave function density (dark green solid), and target state (yellow outline).
			Middle: Graph view. Each solution corresponds to a point in the graph. The duration axis is normalized by an approximate $T^{F=0.99}\st{QSL}$ bound and is divided into 12 blocks. The uppermost green line (the challenge curve) corresponds to the best solution found by conventional methods prior to the launch of the game. Any point can be selected for optimization.
			Bottom: The player can invoke the embedded optimization algorithm (\pgrape) by toggling the switch.
		}

		\label{fig:qm2}

	\end{center}
\end{figure}

The player is faced with two main views within a level: the dynamic view and the graph view. A simplified version of these is shown in Fig.~\ref{fig:qm2}. In the dynamic play view, 
the player generates an initial solution by dragging the round cursor (lower solid turquoise dot) to the round target point (upper transparent turquoise dot). The position of the cursor is linearly mapped by $f_1$ and $f_2$ (Eqs.~\ref{eq:f1f2}) to the instantaneous control function values, $\left\{u_1(t), u_2(t)\right\} = \left\{f_1(x\st{cursor}(t)), f_2(y\st{cursor}(t)\right\})$, which dynamically alter the potential (green). 
The instantaneous wave function density (dark green solid with red line) propagates in the current potential (green line) and the goal is to match the target state density (yellow line) without excess excitation at the end, as the target states are stationary. 
The path traced by the cursor can have any shape, but it is clamped to remain within the control boundaries (turquoise dashed bounding box).
In the graph view, the solution is indicated by a dot on an $F(T)$ graph corresponding to its final fidelity and transfer duration. Note that the fidelity axis is nonlinear and the $T$ axis is normalized by an approximate reference $T^{F=0.99}\st{QSL}$ bound found by a conventional optimization before the launch of the game.
The $T$ axis is divided into 12 blocks of equal size.
Each block contains three green lines of increasing fidelity. 
Points are accrued within each block by placing a solution with the highest fidelity possible relative to these.
The uppermost green line is the \textit{challenge curve}, which are reference results found by conventional \grape methods. 
The remaining lower green lines are motivational game elements without any scientific significance.


In either view the player can click a button to start or stop a \grape optimization of the currently selected solution. During optimization the point will climb in the graph view and the final wave function density will become increasingly similar to the target outline in the dynamic view. 
The player also has other tools available (not shown) such as replaying the time evolution of a solution or setting a solution as a \textit{ghost} that plays along during dynamic play. This can help the player determines strategies by recognizing certain high-level characteristics inspired by the solutions found by the optimizer, or players can try to mimic certain behaviors.

\clearpage
\section{Numerics and Control Problems}
\label{app:numerics}
The state transfers $\psi_0 \rightarrow\psitgt$ considered in this paper are simulated and solved in the QEngine \cite{sorensen2019qengine}, our C++ software package for quantum optimal control. On top of this library is an interfacing code layer exposing functionality and facilitating interoperability with the Unity C\# game code.

The system dynamics is governed by the Schr\"odinger equation $i \hbar \dot \psi = \op H\psi$, where the Hamiltonian is parametrized by up to two control parameters, $\op H = \op H(u_1(t),u_2(t))$. 
More conveniently we can write the state at final time $T$ as
$\ket{\psi_T} =\U\ket{\psi_0}$ where $\U= \mathcal T \exp\left({-i\int_{0}^{T} \op H(u_1(t'),u_2(t')) \mathrm{d}t'} \right)$ is the time-evolution operator and $\mathcal{T}$ denotes time-ordering.
 The goal is thus to find a set of optimal controls $\{u\st{1}^*(t),u\st{2}^*(t)\}$ that implements $\psi_T= \psitgt$ (up to a global phase) or, equivalently, maximizes the fidelity $F=|\braket{\psitgt|\psi_T}|^2 = |\braket{\psitgt| \U|\psi_0}|^2  \leq 1$.

The continuous time parameter is discretized on a regular grid $t \in \{t_1, t_2, \dots, t_j, \dots t_{n_t}\}$ with sufficiently small spacing $\delta t$, and the spatial dimension is similarly discretized. We may then approximate $\U \approx  \prod_{j=1}^{{n_t}-1}\U_{t_{j}} = \U_{t_{{n_t}-1}} \dots\U_{t_{1}}$ where $\U_{t_{j}} = \U(u_1(t_j), u_2(t_j))$ is the local time evolution operator, $\psi_{t_{j+1}}= \U_{t_{j}} \psi_{t_j}$.
The control functions are bounded 
\begin{align}
\{u_1\ut{min}, u_2\ut{min}\} \leq \{u_1(t), u_2(t)\}  \leq \{u_1\ut{max}, u_2\ut{max}\}, \label{eq:controlbounds}
\end{align}
and the control values fixed at $t=0$ and $t=T$, depending on the level. 
For optimizations only, controls and bounds are linearly transformed into  $0 \leq u_i(t)\leq 1$ for $i=1,2$. 

\textit{Units}:
For numerical purposes we obtain non-dimensionalized \cite{sorensen2019qengine} working equations such that effectively $\hbar= m = 1$ and
\begin{align}
i \dot \psi = -\kappa \frac{\partial^2\psi}{\partial x^2} + V\psi + g |\psi|^2 \psi, \label{eq:nondim}
\end{align}
where $\kappa$ is a constant that can be used to gauge the units. 
SI and simulation units are related by $\alpha\st{SI} = \mu\st{[\alpha]} \alpha\st{sim}$ where $\mu\st{[\alpha]}$ is the chosen unit for the dimension of quantity $\alpha$ and $\alpha\st{sim}$ is the dimensionless number entering e.g. Eq.~\eqref{eq:nondim} (the subscript is henceforth omitted for brevity and quantities written without units imply simulation values). 
We take the atomic species to be rubidium-87 atoms such that the unit of mass is $\mu\st{mass} = m\st{Rb} = 87\,\si{amu}$ and we take the energy unit to be $\mu\st{energy} = \hbar/ \mu\st{time}$.
Fixing two elements of the triplet $\{\kappa, \mu\st{length}, \mu\st{time}\}$ 
determines the remaining element to produce Eq.~\eqref{eq:nondim}. 
\begin{itemize}[itemindent=\theitemindent em,align=left,  leftmargin=*]
	\item[\textit{Bring Home Water}:] The units are fixed by
	\begin{subequations}
		\begin{align}
		\kappa &= 0.5, \quad
		\mu\st{length} = 532\,\si{nm},\\
		\Rightarrow		\mu\st{time} & = 2 \mu\st{mass} \kappa \mu\st{length}^2/\hbar = 0.38731\,\si{ms}.
		\end{align}
	\end{subequations}
	Time steps are of size $\delta t = 3.5\cdot 10^{-4}$ and $x \in [-3, 3]$ with $n_x = 256$ grid points. 
	The sum of the tweezer potentials, initial control values, and bounds are given by
	\begin{subequations}
		\begin{align}
		V(u_1,u_2) =\; &u_2(t) \cdot \exp\bigg({\frac{-2(x-u_1(t))^2}{\sigma^2}} \bigg) \notag \\
					        +&A\cdot \exp\bigg({\frac{-2(x-x_0)^2}{\sigma^2}}  \bigg), \\
		\{u_1(0), u_2(0)\} &= \{u_1(T), u_2(T)\} = \{-1, -130\} \\
		\{-2, -150\} &\leq \{u_1(t), \hspace{1.1mm}u_2(t)\} \hspace{1.1mm} \leq \{2, 0\}	        
		\end{align}
	\end{subequations}
	where $\sigma = 0.5$, $x_0 = 1$, $A = -130\; (\approx -53.42\,\si{kHz} \cdot h)$. 

	The frequency of the harmonic approximation when the control tweezer is maximally deep and centered on $x_0$ is $\omega = \sqrt{-4(A + u_2\ut{max})/ (m \st{Rb} \sigma^2)} \approx 66.93\; (\approx 436.8 \;\si{kHz})$.
	
	\item[\textit{Shake Up}:] The units are fixed by
	\begin{subequations}
		\begin{align}
		\mu\st{length} &= 1\,\mu\si{m}, \quad 	\mu\st{time}  = 1\,\si{ms}, \label{eq:units1}\\
		\Rightarrow	\kappa &= \hbar \mu\st{time}/(2\mu\st{mass} \cdot \mu\st{length}^2) = 0.36537. \label{eq:units2}
		\end{align}
	\end{subequations}
	Time steps are of size $\delta t = 1\cdot 10^{-3}$ and $x \in [-2, 2]$ with $n_x = 256$ grid points. 
	The atom chip potential (see Ref.~\cite{sorensen2019qengine} and references therein), initial control values, and bounds are given by
	\begin{subequations}
		\begin{align}
		V(u_1) &= \sum_{r=2,4,6} p_r \cdot (x-u_1(t))^r \\
			u_1(0) &=  u_1(T) = 0, \hspace{0.5cm} -1 \leq u_1(t) \leq 1
		\end{align}
	\end{subequations}
	with coefficients and non-linear coupling strength 
	\begin{align} 
	p_2 &= \hspace{0.25cm}65.8392, && p_4 \hspace{0.2cm}= 97.6349,  \\
	p_6 &= −15.3850, &&g\st{1D}= 1.8299. 
	\end{align}

	\item[\textit{Splitting}:] 	The units are the same as in Eqs.~\eqref{eq:units1}-\eqref{eq:units2} with the addition of $\mu\st{magnetic} = 1\,\si{G}= 10^{-4}\;\si{T}$. 
	Time steps are of size $\delta t = 1\cdot 10^{-3}$ and $x \in [-3.5, 3.5]$ with $n_x = 256$ grid points. 
	The atom chip potential for $\omega = 2\pi\cdot 1.26\,\si{MHz}$ after non-dimensionalization (see also Refs.~\cite{lesanovskyPotential,sorensen2018approaching}), initial control values, and bounds are given by
		\begin{subequations}
			\begin{align}	
			V(u_1) &= p \cdot \sqrt{ \left(B_S(x) - B_\omega\right)^2	+  
			\left(\frac{(0.5 + 0.3 u_2(t))}{2 B_S(x)} B_I\right)^2}, \\
			u_2(0) &=  0, \hspace{0.5cm} u_2(T) = 1, 
			\hspace{0.5cm} 
			 0 \leq u_2(t) \leq 1,
			\end{align}
		\end{subequations}
		In this form, $p=(\mu\st{magnetic} \cdot \mu\ut{Bohr} m_F g_F )/ \mu\st{energy} = 8794.1$ is an overall factor ($\mu\st{magnetic}$ has been factored out from under the square root), and $m_F = 2$, $g_F = 1/2$ are the internal hyperfine state and Landé factors, respectively.
		Additionally, $B_\omega = 0.9$ and $B_S (x)= \sqrt{(\mathrm{Gr} \cdot  x) ^2 + B_I^2}$, where $\mathrm {Gr} = 0.2$ is a magnetic field gradient and $B_I = 1$. We take $g\st{1D}= 1.8299$.

\end{itemize}

\section{Algorithms and Resources}
\label{app:algorithms}
Here we expand on the algorithms described in Sec.~\ref{sec:algseed} and the optimization resources employed in Secs.~\ref{sec:bhw}-\ref{sec:shakeup}.

Outside the game, optimizations were performed in large batches on a computer cluster. Each seed was optimized until it met any of the pertinent termination criteria listed in the following subsections. 
The initial minimal allotted time was roughly $13$ minutes per seed. 
To maximize resource use, excess time for seeds that terminated due to the other criteria was divided equally amongst the remaining seeds and added to their minimal times.

Inside the game,  \pgrape ran locally on the player devices until manually stopped by the player, convergence (step size below $10^{-7}$), or $F=0.999$, incurring no computational cost for us.

\subsection{GRAPE}
Our variant of the standard \grape methodology includes derivative regularization ($\gamma = 10^{-6}$) and boundary  cost terms ($\sigma = 2\cdot 10^3$, see \cite{sorensen2019qengine} for details). 

The pertinent stopping conditions for \grape are:  exceeding the allotted optimization wall time (minimum $\sim 13$ minutes), exceeding the fidelity threshold ($F\geq 0.999$), or subceeding the line search threshold ($\alpha^{k} < 10^{-7}$). 
In the case of \pgrape, the wall time condition is replaced by an in-game button as described in the text.

A single \grape iteration requires calculating the gradient ($2 n_t$ time steps), calculating the \textsc{l-bfgs} quasi-Newton search direction (negligible cost), and performing a line search ($\sim5 n_t$ time steps on average in our implementation), for a total time step cost of $7n_t$.
 This algorithm is fully exploitative (local) with an exploratory (global) component induced through multi-starting \cite{ugray2007multistart}. 

\subsection{(Discrete) Stochastic Ascent}
This section reviews and expands the analysis of the stochastic ascent algorithm from Ref.~\cite{sels2018stochastic}.
In the following, we assume a single control parameter for simplicity.
The time axis is segmented into $n_b$ bins of equal width $w$ which have the same control value, such that $n_t = w \cdot n_b$. For example for $w=3$
$$\vec u = \{ \underbracket[0.5pt]{u_{t_1}, u_{t_2}, u_{t_3}}_{u_{b_1}} ,  \underbracket[0.5pt]{u_{t_4}, u_{t_5}, u_{t_6}}_{u_{b_2}} , \dots,  \underbracket[0.5pt]{ u_{t_{n_t-2}},u_{t_{n_t-1}}, u_{n_t}}_{u_{b_{n_b}}} \}. $$
The propagator for the first bin is $\U_{b_1} = \U_{t_3} \U_{t_2} \U_{t_1}$ and the pattern continues for the other bins
\footnote{Calculations could in principle be sped up by subsuming the individual propagators into a single step of length $w \cdot \delta t$, although such approach would require small enough $w$ to ensure the error associated with time evolution remains small.}.
We then allow $u_{b_k}$ to assume only values from a predefined discrete set $\Omega = \{u_d\}_{d=1}^{n_d}$
(this choice is discussed at the end of the section). 
Using $n_d=128$, these values are linearly spaced from the lower to the upper control boundary, see \eqref{eq:controlbounds}.

Updating $u_{b_k}$ is done by exhaustively computing the fidelity for all possible values in $\Omega$ and setting $u_{b_k}$ corresponding to the maximal value,  
$$u_{b_k} \leftarrow \underset{u_{b_k} \in \Omega}{\mathrm{argmax}} \; F(u_{b_1}, \dots, u_{b_{k-1}},u_{b_k}, u_{b_{k+1}},\dots,u_{b_{n_b}}),$$
while keeping the other control values fixed (discrete coordinate ascent \cite{wright2015coordinate}).
When $u_{b_k}$ has been updated, it is not chosen for further updates until all the remaining points have also been updated. 
Updating all points once constitutes an iteration and the sequential control update order is stochastic within each of these. 
As is, bandwidth limitations are only imposed by the choice of $n_b$, but one could easily accommodate a derivative regularization term as in \grape.
The pertinent stopping conditions are: exceeding the allotted optimization wall time (minimum $\sim 13$ minutes), exceeding the fidelity threshold $(F\geq 0.999)$, or when the algorithm achieves no gain in fidelity by changing any of the control values. 

Exhaustive evaluation of the fidelity for bin $k$ can be sped up for the linear Schr\"odinger equation ($g=0$), 
\begin{align} 
F(u_{b_k}) = |\Braket{\psitgt| \psi_{t_{n_t}}}|^2 = |\Braket{\chi_{b_{k+1}}| \U_{b_{k}} |\psi_{b_{k}}}|^2
\label{eq:fidelity}
\end{align}
since the forward-propagated $\ket{\psi_{b_{k}}} =  \prod_{j=1}^{k-1} \U_{b_{j}} \ket{\psi_0}$ and backward-propagated $\ket{\chi_{b_{k+1}}} =  \prod_{j=n_b}^{k+1} \U^\dagger_{b_{j}} \ket{\psitgt}$ vectors for all times up to and after bin $k$, respectively, only need to be calculated and cached once per bin update. 
Upon finishing the evaluation, the update is applied to the control and forward-propagated state.
Additionally, the matrix representation of the time-evolution operators $\U_d$ corresponding to every element in $\Omega$ can be precomputed and cached in memory, changing the time-stepping method to a single matrix-vector multiplication \cite{sels2018stochastic} instead of the Fourier split-step method.
 
The first control value update requires $(n_t + n_d)$ time steps after which the forward/backward vectors have been initialized and cached. 
Calculating the new forward/backward vectors and updating for a subsequent bin at $k'$ requires only $w |k-k'|$ time steps when re-using the old vectors. 
If $k < k'$ the $\psi$ cache is updated and the $\chi$ cache otherwise. 

 We may write the average time step distance $w \braket{|k-k'|} = w \rho  n_b = \rho  n_t$ where $\rho \approx 1/3$ is found empirically. Performing subsequent updates thus costs $(n_t/3  + n_d)$ time steps when averaged over all bins.
The average number of time steps required to complete a full iteration is thus $n_b( n_t/3 + n_d)$,
except the first iteration which costs an additional $2n_t/3 $ due to forward/backward vector cache initialization \footnote{$(n_t+n_d) + (n_b - 1)(n_t/3 + n_d) = \frac{2 }{3}n_t + n_b( n_t/3 + n_d)$}.

In the non-linear case $g\neq 0$, the explicit state dependence has severe consequences for the algorithm's feasibility.
First, the time evolution operators $\U_j$ cannot be precomputed since they depend on $\psi$, which changes as the control changes.
Second, it does not make sense to maintain a cache for backward-propagated vectors; altering the control at $t_k$ changes the $\psi$ state trajectory from $k$ to $n_t$ and the backward-propagated vectors depend on these in the non-linear case. In this case one may just as well evaluate the fidelity at $t_{n_{t}}$ using the first equality in Eq.~\eqref{eq:fidelity}.
The $\psi$ cache only needs to be updated when $k < k'$, yielding $w \braket{|k-k'|}_{k < k'} = w n_b/6 = n_t/6$. 
From there, evaluating the fidelity of a single $u_{b_k} \in \Omega$ requires $w |n_b-k|$ time steps which must be done for all $n_d$ elements. 
Averaging over a full iteration yields $w\Braket{|n_b- k|} = w n_b/2 =  n_t/2$. Consequently, the number of time steps needed to update a single point $u_{b_k}$ changes as
$n_t/3 \rightarrow n_t/6$ and
 $n_d \rightarrow  n_d n_t  /2$.

\begin{table}
\begin{center}
	\begin{tabular}{m{2.4cm} | m{2.6cm} m{3cm}}
		& Linear $(g=0)$& Non-Linear $(g\neq0)$\\ 
		\hline 
		\vspace{0.2cm}
		Caching & $\U_d, \psi, \chi$ &  $\psi$\\  
		Time evolution & Matrix $\times$ vector & Fourier split-step\\
		Time steps/iter  & $n_b(\frac{1}{3} n_t + n_d)$ & $n_b(\frac{1}{6} n_t + \frac{1}{2}n_t n_d)$  \\ 
	\end{tabular} 
\end{center}
	\caption{Summary for $g=0$ and $g\neq 0$ for \sa.}
	\label{tab:sa}
\end{table}

\noindent Thus roughly an additional $n_b n_d n_t/2$ time steps must be performed when $g\neq 0$, each of which has an increased computation time because $\U_d$ cannot be cached. 
The differences between the two cases is summarized in Table~\ref{tab:sa}.

The speed with which the stochastic (coordinate) ascent operates comes at the cost of not being able to perform correlated, simultaneous control value updates between bins. This could easily be remedied, in principle, by updating $n_p$ bins instead of just a single one. However, such an approach is untenable for the discrete version even for small $n_p > 1$ if one desires exhaustive search: the update cost would then depend on the largest index distance between the parameters and an exponential number of discrete combinations $n_d^{n_p}$. 

The fast linear evaluation ($g=0$) for the \sa algorithm 
is independent of the choice to restrict control values to a discretized set of values. For example, it would be straightforward to use the same fast evaluation methodology
with derivative-based methods and perform line searches. 
Such a change in update rule shifts the exploration vs. exploitation trade-off: the discrete version is in a sense fully exploratory (the search is globally exhaustive), but only along one axis at a time. Obviously, discretization and fixing the remaining axes produces a reduced representation of the underlying control landscape with respect to which the discrete version exhibits a mix of global- and local search properties.
Abandoning discretization and performing line searches turns the algorithm into a fully exploitative one, but again only along a single axis. 
This is the more standard version of coordinate ascent \cite{nocedal2006numerical}.
In our implementation, line searching usually requires about $5 \ll n_d$ objective evaluations \footnote{The discrete \sa algorithm could also be adapted to e.g. only check an informed subset of the $n_d$ possibilities in $\Omega$, or let $\Omega$ itself be dynamically changing}, allowing potentially orders of magnitude fewer time steps ($n_d \rightarrow 5$ in Table~\ref{tab:sa}) when close to an optimum.
Although such an approach does not allow caching of the unitary time evolution operator since the controls can take any value, the aforementioned benefits should more than compensate for this during the local adjustment phase.
In this setting, however, the advantageous convergence rates associated with derivatives and adaptive step sizes is only with respect to the chosen axis -- there are no theoretical guarantees for convergence in the full dimensional landscape \cite{nocedal2006numerical}. 
In light of these observations, it would be interesting to combine the three methodologies with handover techniques, for example starting with the most global algorithm and ending with the most local one
\begin{center}
	discrete \sa $\rightarrow $ gradient \sa $\rightarrow$ \grape
\end{center}
The performance of such a combination would be interesting to try on the different seeding strategies discussed in the main text and is a potential subject of future work.
%

%

\bibliography{/Users/au446513/projects/bibmaster/references.bib}

\end{document}